\documentstyle[epsfig,aps,eqsecnum,preprint,tighten]{revtex}

\begin{document}

\preprint{\vbox{
\hbox{TUM/T39-98-10}
\hbox{SNUTP-98-037}
\hbox{hep-ph/9901354}
}}

\draft

\title{Baryon masses in large $N_c$ chiral perturbation theory%
\footnote{Work supported in part by BMBF}}

\author{Yongseok Oh%
\footnote{Alexander von Humboldt fellow,
E-mail address: {\tt yoh@phya.snu.ac.kr}}}

\address{
  Physik-Department,
  Technische Universit\"at M\"unchen,
  D-85747 Garching, Germany \\
and
  Research Institute for Basic Sciences and Department of
  Physics, Seoul National University, Seoul 151-742, Korea}

\author{Wolfram Weise}

\address{
  Physik-Department,
  Technische Universit\"at M\"unchen,
  D-85747 Garching, Germany}

\maketitle

\begin{abstract}

We analyse the baryon mass spectrum in a framework which
combines the $1/N_c$ expansion with chiral perturbation theory.
Meson loop contributions involving the full SU(3) octet of pseudoscalar
Goldstone bosons are evaluated, and the influence of explicit chiral and
flavor symmetry breaking by non-zero and unequal quark masses is
investigated. We also discuss sigma terms and the strangeness contribution
to the nucleon mass.

\end{abstract}



\section{Introduction}

The large $N_c$ limit, where $N_c$ is the number of colors, is a useful
device to understand many systematic features of baryon properties
\cite{tHoo74a,Wit79b}, such as the $1/N_c$ scaling of various physical
quantities.
In a series of papers, Dashen and Manohar \cite{DM93a,DM93b}, and
Jenkins \cite{Jen93b} have discussed the $1/N_c$ structure of baryon
properties, and the framework for combining chiral symmetry with the
large $N_c$ aspects of QCD has been developed by many authors
\cite{DJM94,JL95,Jen96,LM94,Luty95,BL96,Coh95}.
In Refs. \cite{JL95,Luty95}, the baryon octet and decuplet mass spectra
were discussed in this framework and the baryon mass relations were
derived.
However, although those works successfully reproduce mass relations at 
tree level, they do not compute all possible terms allowed by the chiral
and large $N_c$ expansions.

The baryon mass spectrum was re-examined in conventional baryon chiral
perturbation theory by Borasoy and Meissner \cite{BM96a,BM97}.
To compute the baryon masses to order $m_q^2$,
where $m_q$ is the quark mass, the decuplet degrees of freedom are
integrated out to give counter terms, and some low-energy constants
are determined from resonance saturation.
However, when we work with the $1/N_c$ expansion, the octet and
decuplet states are degenerate at the leading order, and the decuplet
fields must be treated explicitly.

In this paper, we re-examine the baryon masses in chiral perturbation
theory taking into account the $1/N_c$ counting based on the techniques
developed in the literature, e.g., in Refs. \cite{LM94,Luty95,BL96}.
This enables us to investigate the $1/N_c$ structure of the baryon
properties and the meson-baryon interactions in a systematic way.
The baryon axial current matrix elements and the strangeness contribution
to the nucleon mass are computed as well.
Some of these topics were discussed in the literature
\cite{JL95,Luty95,BL96} focusing on the leading order terms in $1/N_c$
expansion (up to one loop corrections).
In this paper, we perform the calculations to the next orders and
discuss a difficulty which arises in computing the one loop corrections in
a way which is consistent with the $1/N_c$ expansion.
This paper is organized as follows.
In the next Section, we briefly discuss the formalism of this approach.
In Section III, we compute the baryon axial current up to one-loop
corrections. The baryon mass formulas are given in Section IV.
The one-loop corrections to the baryon masses are calculated in Section V.
We discuss the strangeness contribution to the nucleon mass and the sigma
term in Section VI and then finish with a summary and conclusion in
Section VII.
Explicit expressions of baryon wave functions and some detailed formulas
are given in Appendices.

\section{Formalism}

We start with a brief review of the construction of baryon states in the
large $N_c$ limit, referring to Refs.\cite{LM94,Luty95,JL} for further
details.
The baryon states with $N_c$ quarks can be written as follows:
\begin{equation}
| B \rangle \equiv {\cal B}^{a_1 \alpha_1 \dots a_{N_c} \alpha_{N_c}}
\varepsilon^{A_1 \dots A_{N_c}} q^\dagger_{a_1 \alpha_1 A_1} \dots
q^\dagger_{a_{N_c} \alpha_{N_c} A_{N_c}} | 0 ),
\end{equation}
in particle number space, where $a_i$ are the flavor indices, $\alpha_i$
the spin indices and $A_i$ the color indices.
The quark creation and annihilation operators $q^\dagger$ and $q$ satisfy
the usual anti-commutator relations for fermions.
The symmetric tensor ${\cal B}$ is characteristic of each given baryon wave
function.
Since the baryons are in color-singlet states, however, it is more
convenient to work with
\begin{equation}
| B ) \equiv {\cal B}^{a_1 \alpha_1 \dots a_{N_c} \alpha_{N_c}}
\alpha^\dagger_{a_1 \alpha_1} \dots \alpha^\dagger_{a_{N_c} \alpha_{N_c}}
| 0 ),
\end{equation}
by dropping the explicit color indices, where the operators
$\alpha^\dagger$ and $\alpha$ are {\em bosonic} operators satisfying
the usual commutator relations.
For short-hand notation, we label the quark operators as
\begin{eqnarray}
&&
\alpha_1 \equiv \alpha_{u\uparrow}, \qquad
\alpha_2 \equiv \alpha_{u\downarrow}, \qquad
\alpha_3 \equiv \alpha_{d\uparrow}, \nonumber \\
&&
\alpha_4 \equiv \alpha_{d\downarrow}, \qquad
\alpha_5 \equiv \alpha_{s\uparrow}, \qquad
\alpha_6 \equiv \alpha_{s\downarrow},
\end{eqnarray}
so that $\alpha_1^\dagger$ creates $u$-quark with spin-up, and so forth.

There is an ambiguity when we extrapolate the physical baryon states
to large $N_c$.
As in the literature, we keep the spin, isospin and
strangeness of baryons as $O(1)$ in the large $N_c$ limit.
For example, the nucleon state in large $N_c$ limit has spin 1/2,
isospin 1/2 and no strangeness.
This can be done by acting with spin-flavor singlet operators on the
physical baryon states.
For example, the proton spin-up state can be written as
\begin{equation}
|p, +{\textstyle\frac12}) = C_N \alpha_1^\dagger (A_s^\dagger)^n |0),
\end{equation}
where $n = (N_c-1)/2$ and $C_N$ is the normalization constant.
The spin-isospin singlet operator $A_s^\dagger$ is defined as
\begin{equation}
A_s^\dagger = \alpha_1^\dagger \alpha_4^\dagger - \alpha_2^\dagger
\alpha_3^\dagger.
\end{equation}
One can easily verify that this state reduces to the usual quark model
state in the real world with $N_c=3$.
The complete list of the baryon octet and decuplet states can be found
in Appendix A.

Next we define a one-body operator $\{ X \}$ in spin-flavor space as
\begin{equation}
\{ X \} = \alpha_{a\alpha}^\dagger X_{ab}^{\alpha\beta} \alpha_{b\beta},
\end{equation}
so that its expectation value on baryon states is at most of $O(N_c)$.
In a similar way, one can define 2-body operators $\{ X \} \{ Y \}$ and
3-body operators $\{ X \} \{ Y \} \{ Z \}$, and so on.
Then, it is found that the coefficient of an $r$-body operator is at
most $O(1/N_c^{r-\ell-1})$, where $\ell$ is the number of inner quark
loops \cite{LM94,CGO94}.
This enables us to treat the coupling constants as $O(1)$ quantities in
the large $N_c$ expansion by writing the $N_c$-dependence of the operators
explicitly.

By direct evaluation, one can see the well-known commutator relation,
\begin{equation}
[ \{ X \}, \{ Y \} ] = \{ [ X, Y ] \}.
\label{Comm}
\end{equation}
Note that the left-hand side is naively a two body operator whose
expectation values can be of $O(N_c^2)$, whereas the right-hand
side is a one-body operator whose expectation values are of $O(N_c)$
at most.
This means that the order of an operator in $1/N_c$ counting reduces when
we have a commutator structure as in Eq. (\ref{Comm}).
This plays an important role in the large $N_c$ analyses of the baryon
properties.

We will discuss the explicit forms of some operators which appear in
the calculation of baryon axial currents and masses in the next Sections.

\section{Baryon axial currents}

\subsection{Tree level}

Our starting point is the chiral meson-baryon effective Lagrangian.
Baryon matrix elements of this Lagrangian involve the meson-baryon
interaction in the following form:
\begin{eqnarray}
\langle {\cal L}_{\rm eff} \rangle &=& g \, (B| \{ A^\mu \sigma_\mu \} |B)
+ \frac{h}{N_c} \, (B| \{ A^\mu \} \{ \sigma_\mu \} |B)
+ \dots,
\label{Lag:mb}
\end{eqnarray}
where $\sigma^\mu$ is the baryon spin matrix,%
\footnote{In the baryon rest frame, $\sigma^\mu=(0,\bbox{\sigma})$ with
the usual Pauli matrices $\sigma^i$.}
and the dots denote higher order terms.
The axial field $A_\mu$ is defined as
\begin{eqnarray}
A_\mu &=& \frac{i}{2} ( \xi \partial_\mu \xi^\dagger
- \xi^\dagger \partial_\mu \xi ),
\end{eqnarray}
where $\xi = \exp ( i \Pi / f)$ with the meson decay constant $f$.
The SU(3) matrix field $\Pi$ represents the octet of pseudoscalar
Goldstone bosons.
It is defined as
\begin{equation}
\Pi = \frac{1}{2} \lambda_a \pi_a,
\end{equation}
with the usual Gell-Mann matrices $\lambda_a$ $(a=1, \dots, 8)$.
In Eq. (\ref{Lag:mb}), the $N_c$ factors of operators are given
explicitly, and the coupling constants $g$ and $h$ are of $O(1)$
in the $1/N_c$ counting.

Then the baryon axial current $J_{5,\mu}^a$ reads
\begin{equation}
J_{5,\mu}^a = \frac{g}{2} \, \{ \tilde{T}^a \sigma_\mu \}
+ \frac{h}{2N_c} \, \{ \tilde{T}^a \} \{ \sigma_\mu \},
\end{equation}
from the Lagrangian (\ref{Lag:mb}) with
\begin{equation}
\tilde{T}^a = \frac12 ( \xi \lambda^a \xi^\dagger + \xi^\dagger \lambda^a \xi).
\end{equation}
This gives its matrix elements as
\begin{equation}
(B'| J_{5,\mu}^a |B) = \alpha_{B'B}^a \bar u_{B'} (\sigma_\mu) u_B.
\end{equation}
where $u_B$ is the Dirac spinor of the baryon and
\begin{equation}
\alpha_{B'B}^a = g (B'| \{ {\textstyle\frac12} \lambda^a \sigma^3 \} |B)
+ \frac{h}{N_c} (B'| \{ {\textstyle\frac12} \lambda^a \} \{ \sigma^3 \} |B),
\end{equation}
at the tree level.

By using the wave functions given in Appendix A, we can compute the
baryon axial current straightforwardly.
A naive investigation of each term gives that,
despite the $1/N_c$ factor, the $h$ term contribution is expected to be
of the same order as that coming from the $g$ term.
This is because the $h$ term contains a 2-body operator whose expectation
value can be $O(N_c^2)$, thus the leading order of the $h$ term
contribution can be of $N_c$.
However, close inspection shows that the $g$ term contribution dominates,
because the $h$ term contains the operator $\{ \sigma_\mu \}$ and
our baryon wave functions satisfy $\{ \sigma_\mu \} \sim O(1)$.

The explicit forms of the relevant operators are
\begin{eqnarray}
\{ {\textstyle\frac12} \lambda^{1+i2} \sigma^3 \} =
\alpha_1^\dagger \alpha_3 - \alpha_2^\dagger \alpha_4, &\qquad&
\{ {\textstyle\frac12} \lambda^{4+i5} \sigma^3 \} =
\alpha_1^\dagger \alpha_5 - \alpha_2^\dagger \alpha_6,
\nonumber \\
\{ {\textstyle\frac12} \lambda^{1+i2} \} \{ \sigma^3 \} =
\alpha_1^\dagger \alpha_3 + \alpha_2^\dagger \alpha_4, &\qquad&
\{ {\textstyle\frac12} \lambda^{4+i5} \} \{ \sigma^3 \} =
\alpha_1^\dagger \alpha_5 + \alpha_2^\dagger \alpha_6,
\label{opform1}
\end{eqnarray}
so that we obtain
\begin{eqnarray}
\alpha_{pn}^{1+i2} &=& \frac{g}{3} (N_c+2) + \frac{h}{N_c} ,
\nonumber \\
\alpha_{\Lambda \Sigma^-}^{1+i2} &=& \alpha_{\Sigma^+ \Lambda}^{1+i2} =
\frac{g}{3\sqrt2} \sqrt{(N_c-1)(N_c+3)},
\nonumber \\
\alpha_{\Xi^0 \Xi^-}^{1+i2} &=& \frac{N_c g}{9} - \frac{h}{N_c} ,
\nonumber \\
\alpha_{\Sigma^0 \Sigma^-}^{1+i2} &=& -\alpha_{\Sigma^+ \Sigma^0}^{1+i2} =
\frac{g}{3\sqrt2} (N_c+1) + \frac{\sqrt2 h}{N_c} ,
\end{eqnarray}
and
\begin{eqnarray}
\alpha_{p\Lambda}^{4+i5} &=& - \frac{g}{2} \sqrt{N_c+3}
- \frac{h}{2N_c} \sqrt{N_c+3} ,
\nonumber \\
\alpha_{\Lambda \Xi^-}^{4+i5} &=& \frac{\sqrt{N_c}g}{2\sqrt3}
+ \frac{\sqrt3 h}{2N_c} \sqrt{N_c-1} ,
\nonumber \\
\alpha_{p \Sigma^0}^{4+i5} &=& \frac{1}{\sqrt2} \alpha_{n \Sigma^-}^{4+i5}
= \frac{g}{6} \sqrt{N_c-1} - \frac{h}{2N_c} \sqrt{N_c-1} ,
\nonumber \\
\alpha_{\Sigma^0 \Xi^-}^{4+i5} &=&
\frac{1}{\sqrt2} \alpha_{\Sigma^+ \Xi^0}^{4+i5}
= \frac{5g}{6\sqrt3} \sqrt{N_c+3} + \frac{h}{2\sqrt{3}N_c} \sqrt{N_c+3} .
\end{eqnarray}
These results show that the $h$ term contributions are suppressed as
compared to those of the $g$ terms as we discussed above.
We can also find that the leading order of $\alpha^{1+i2}_{BB'}$ is
$O(N_c)$, whereas $\alpha^{4+i5}_{BB'}$, which changes the baryon
strangeness, is $O(\sqrt{N_c})$.
This shows that the strangeness-changing (i.e., $\Delta S \neq 0$)
baryon axial currents are suppressed as compared to the
strangeness-conserving (i.e., $\Delta S = 0$) baryon axial
currents by $O(\sqrt{N_c})$.
This can be understood from Eq. (\ref{opform1}) by noting that the number
of $u$,$d$ quarks in the baryon wave functions is $O(N_c)$ whereas
that of $s$ quark, i.e., strangeness, is $O(N_c^0)$.
For example, in the case of $\alpha_{pn}^{1+i2}$, acting with $\alpha_3$
(or $\alpha_4$) on the baryon state gives the factor $N_c$, and the
inner product of initial and final baryon wave functions with the proper
normalization constants gives $O(1)$, so that $\alpha_{pn}^{1+i2}$ is
$O(N_c)$.
However, for $\alpha_{p\Lambda}^{4+i5}$, the action with $\alpha_5$ (or
$\alpha_6$) gives $O(N_c^0)$ because our baryon wave functions have the
strangeness of $O(N_c^0)$.
Since the normalization constants of nucleon and $\Lambda$ are $O(1/N_c)$
and $O(1/\sqrt{N_c})$, respectively, we have an additional factor $\sqrt{N_c}$
in the calculation of $\alpha_{p\Lambda}^{4+i5}$, which implies that
the order of $\alpha_{p\Lambda}^{4+i5}$ is $O(\sqrt{N_c})$.

Since the contributions of the $h$ term are suppressed as compared to 
those of the
$g$ term by $O(1/N_c^2)$ for the $\Delta S = 0$ axial currents and by
$O(1/N_c)$ for the $\Delta S = 1$ axial currents, we can neglect the
$h$ term up to next to leading order.
At this order, when we fix $N_c=3$, we can recover the quark model relation
\cite{Luty95},
\begin{equation}
D = g, \qquad F = \frac23 g,
\label{d-f-1}
\end{equation}
by comparing with the results of the baryon chiral perturbation theory
\cite{JM91a,JM91b} in addition to the quark model predictions
\begin{equation}
{\cal C} = -2 g, \qquad {\cal H} = -3 g,
\end{equation}
for the octet-decuplet-meson and decuplet-decuplet-meson coupling
constants, ${\cal C}$ and ${\cal H}$, defined as in Ref. \cite{JM92b}.
When we go further in the $1/N_c$ expansion, we must include the $h$ term, and
we get the modified relations,
\begin{equation}
D = g, \qquad F = \frac{2g+h}{3},
\label{d-f-2}
\end{equation}
as found in Ref. \cite{Luty95}.

We also compute $\alpha_{B'B}^8$ by using
\begin{eqnarray}
&& \{ {\textstyle\frac12} \lambda^{8} \sigma^3 \} =
\frac{1}{2\sqrt{3}} ( N_1 - N_2 + N_3 - N_4 - 2N_5 + 2N_6 ),
\nonumber \\
&& \{ {\textstyle\frac12} \lambda^{8} \} \{ \sigma^3 \} =
\frac{1}{2\sqrt{3}} ( N_1 + N_2 + N_3 + N_4 - 2N_5 - 2N_6 ),
\label{alpha8}
\end{eqnarray}
where we have introduced $N_i = \alpha_i^\dagger \alpha_i$.
This leads to
\begin{eqnarray}
\alpha_{pp}^8 &=& \frac{g}{2\sqrt3} + \frac{h}{2\sqrt3} , \nonumber \\
\alpha_{\Lambda\Lambda}^8 &=&
-\frac{g}{\sqrt3} + \frac{h}{2\sqrt{3}} \left( 1 - \frac{3}{N_c} \right),
\nonumber \\
\alpha_{\Sigma\Sigma}^8 &=&
\frac{g}{\sqrt3} + \frac{h}{2\sqrt{3}} \left( 1 - \frac{3}{N_c} \right),
\nonumber \\
\alpha_{\Xi\Xi}^8 &=&
-\frac{\sqrt3 g}{2} + \frac{h}{2\sqrt{3}} \left( 1 - \frac{6}{N_c} \right).
\end{eqnarray}
{}From these results we find that the leading order of $\alpha_{B'B}^8$
is $O(N_c^0)$ and that the $h$ term provides a leading contribution
together with the $g$ term.
This is because the expectation values of $\{ {\textstyle\frac12}
\lambda^{8} \} \{ \sigma^3 \}$ are $O(N_c)$ whereas those of
$\{ {\textstyle\frac12} \lambda^{8} \sigma^3 \}$ are $O(1)$.
So we conclude that in order to get a consistent result on $\alpha_{B'B}^8$,
one should consider $n$-body ($n \geq 3$) operators in general unless their
coupling constants are suppressed.
{}From the fitted values of $D$ and $F$, one can estimate $g = 0.61 \sim 0.8$
together with $h = -0.02 \sim -0.1$, which shows that $h$ is indeed
small, less than 15\% of $g$, but with opposite sign \cite{Luty95}.
Therefore, one should keep in mind the contributions from $n$-body ($n
\geq 3$) operators in the calculation of $\alpha^8_{B'B}$. 
We have a similar situation when we compute the $\eta$-meson loop
corrections to the baryon masses in Section V.

\subsection{One-Loop Corrections}

\begin{figure}
\centering
\epsfig{file=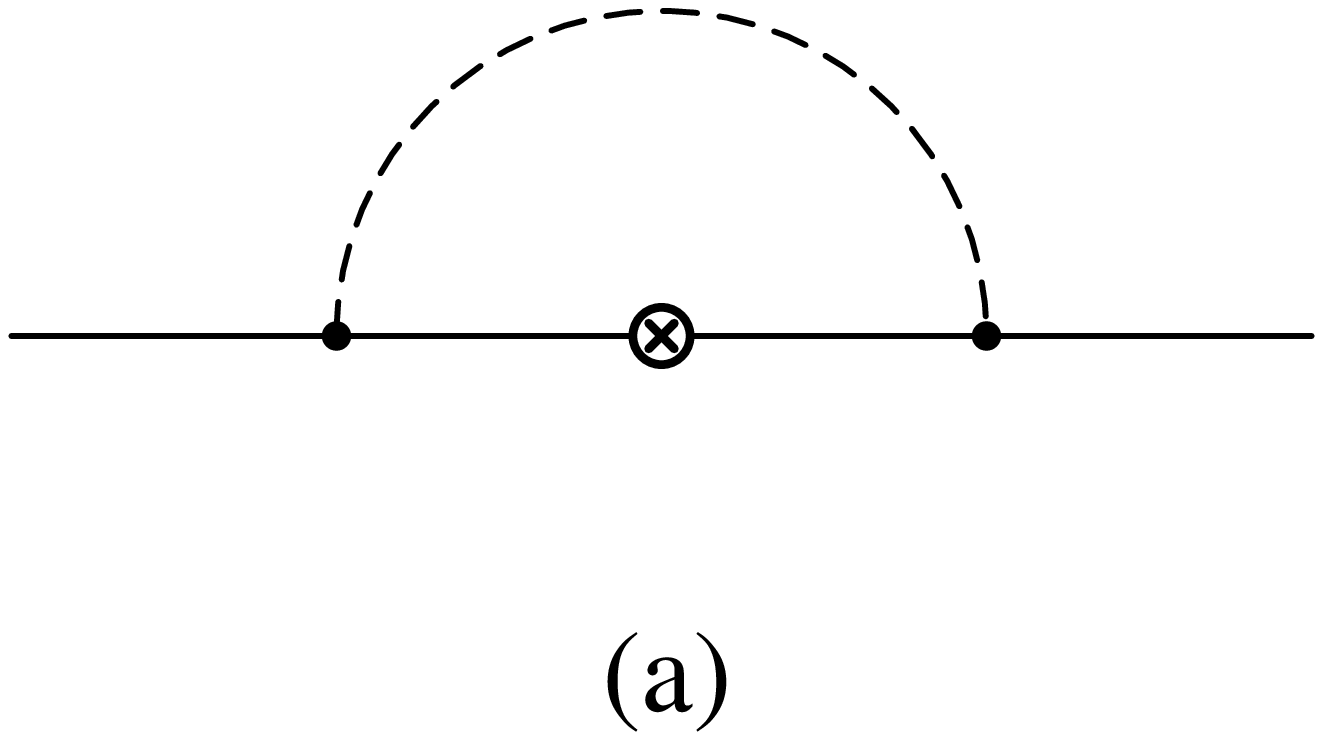, width=0.4\hsize} \qquad
\epsfig{file=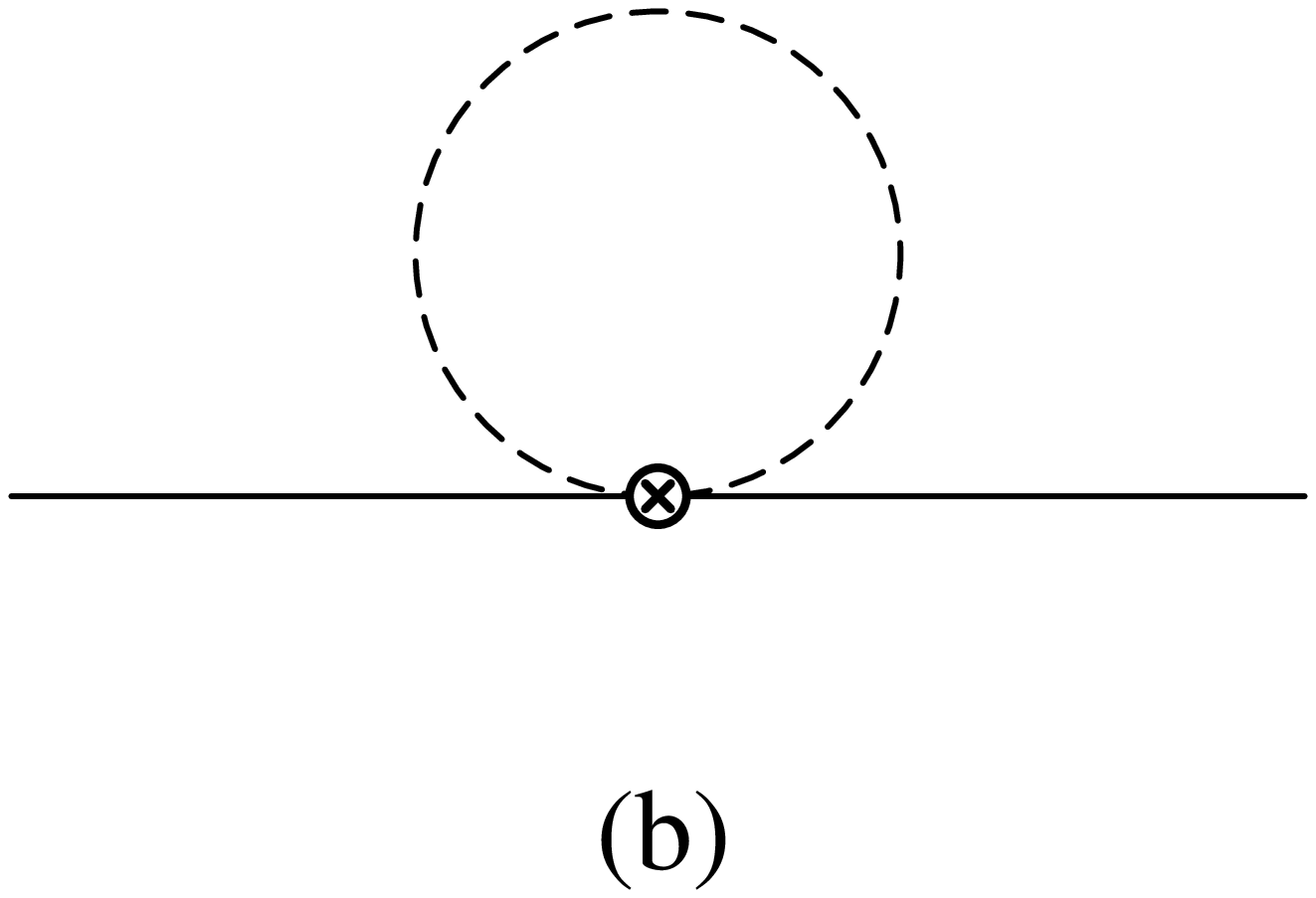, width=0.4\hsize}
\caption{One-loop corrections to the baryon axial current.}
\label{fig:ax1}
\end{figure}

\begin{figure}
\centering
\epsfig{file=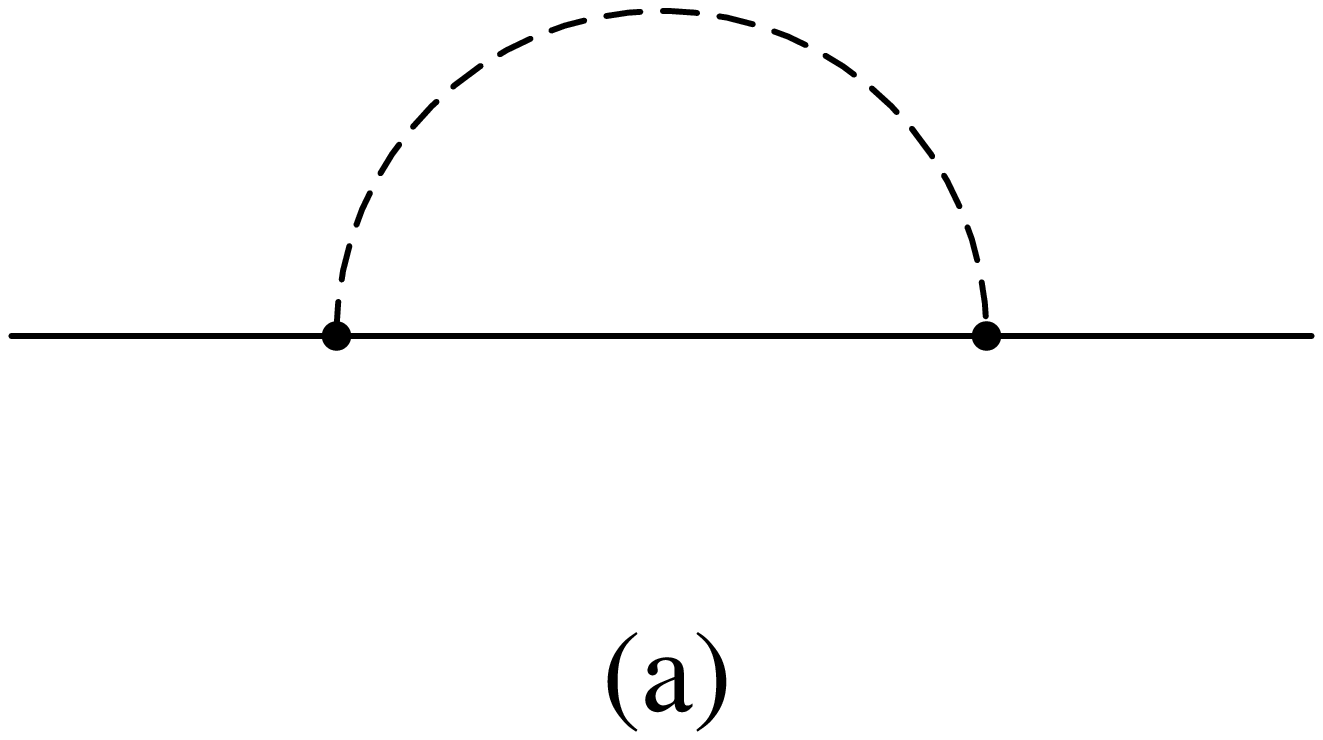, width=0.4\hsize} \qquad
\epsfig{file=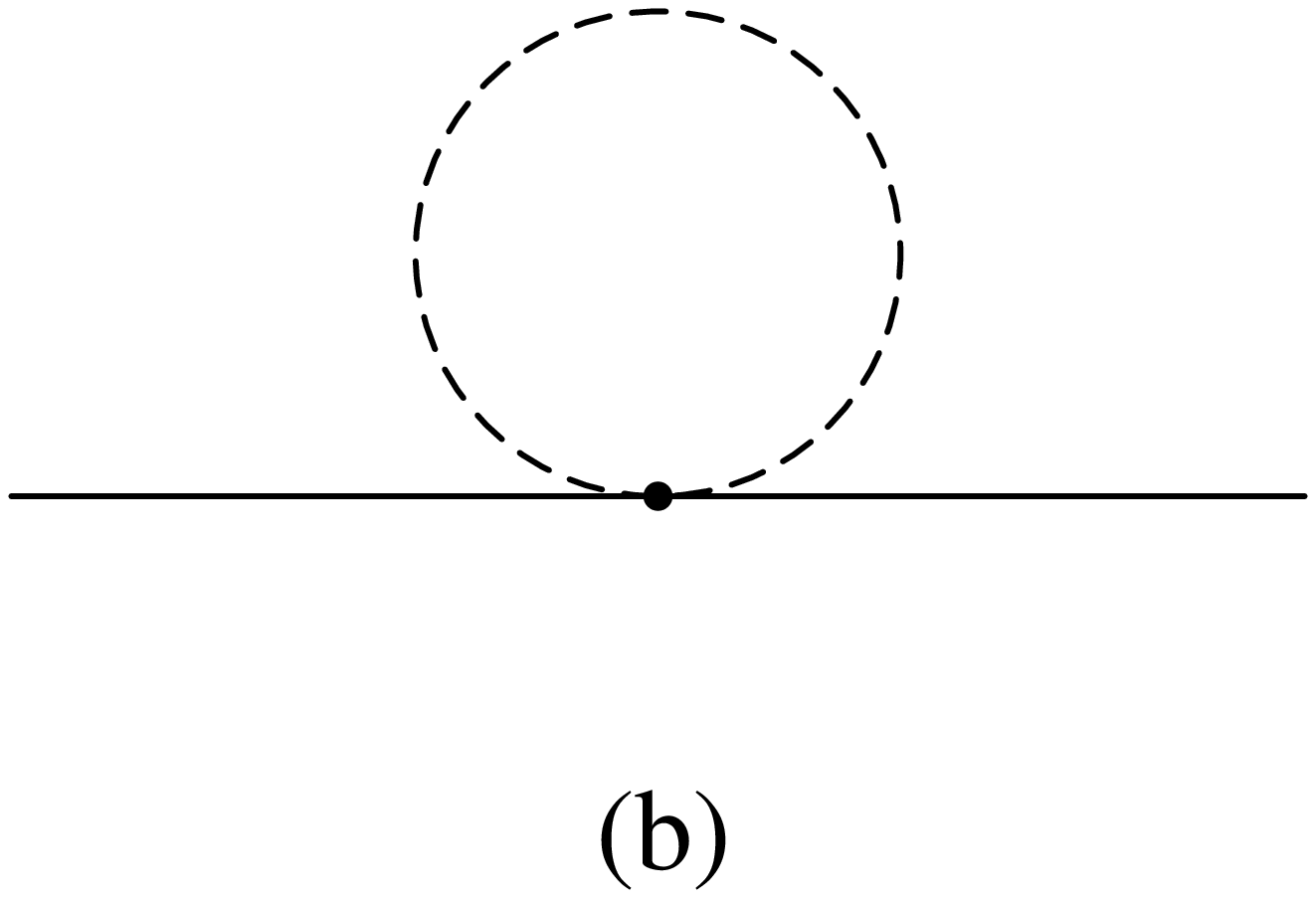, width=0.4\hsize}
\caption{Wave function renormalization of one-loop.}
\label{fig:wf1}
\end{figure}

The one-loop corrections to the baryon axial current in large $N_c$ chiral
perturbation theory as shown in Fig. \ref{fig:ax1} have been discussed in
Refs. \cite{DM93a,DM93b,DJM94}. 
Naively, these loop corrections as they stand are not consistent with the 
$1/N_c$ expansion.
{}From the meson--baryon interactions (\ref{Lag:mb}), each vertex is
related to an operator $X^{ia}$ defined as
\begin{equation}
X^{ia} = g \{ {\textstyle\frac12} \lambda^a \sigma^i \} +
\frac{h}{N_c} \{ {\textstyle\frac12} \lambda^a \} \{ \sigma^i \},
\label{Xia}
\end{equation}
with spin index $i$ and flavor index $a$.
This shows that the meson-baryon coupling is of order $N_c$.
Because of the presence of $1/f$ which scales as $1/\sqrt{N_c}$,
each vertex has a factor $\sqrt{N_c}$.
Then from Fig. \ref{fig:ax1}(a), it is evident that the one-loop
correction is $O(N_c^{2})$ when $\alpha_{B'B}$ is of order $N_c$.
Thus the loop correction dominates the tree level.
However, we have to consider the wave function renormalization terms
(Fig. \ref{fig:wf1}) in the loop calculation.
When combined with Fig. \ref{fig:ax1}, this gives the commutator
structure to the baryon axial current operators, which implies that
the actual order of the one-loop correction is $O(N_c^0)$ when
$\alpha_{B'B}$ is $O(N_c)$.
This suppression was proved up to 2-loop order in Ref. \cite{Kea96}
which concludes that the 2-loop corrections are suppressed
by $1/N_c^2$ as compared to the tree level values.

Explicitly, the one-loop correction to the baryon axial current from
Fig. \ref{fig:ax1}(a) is given by the following expression:
\begin{equation}
V_{B'B}^{ia} = \frac{-i}{f^2} \int \frac{d^4 k}{(2\pi)^4}
\frac{1}{(k \cdot v)^2} \frac{k_\mu k_\nu}{m_{bb'}^2 - k^2}
(B'| X^{\mu b} X^{ia} X^{\nu b'} |B),
\end{equation}
where $m_{bb'} (=m_\pi,m_K,m_\eta)$ is the meson mass in the loop. 
When combined with the wave function renormalization factor $Z_B$ from
Fig. \ref{fig:wf1},
\begin{equation}
Z_B = 1 + \frac{i}{f^2} \int \frac{d^4 k}{(2\pi)^4}
\frac{1}{(k \cdot v)^2} \frac{k_\mu k_\nu}{m_{bb'}^2 - k^2}
(B| X^{\mu b} X^{\nu b'} |B),
\end{equation}
this gives the renormalized baryon axial current operator in the form of
\begin{equation}
X^{ia} + \frac{1}{2f^2} I_{\mu\nu}^{bb'} [ X^{\mu b},
[ X^{ia}, X^{\nu b'} ] ],
\end{equation}
where
\begin{equation}
I_{\mu\nu}^{ab} = -i \int \frac{d^4 k}{(2\pi)^4}
\frac{1}{(k \cdot v)^2} \frac{k_\mu k_\nu}{m_{ab}^2 - k^2}.
\end{equation}

Finally, we get the one-loop correction to the baryon axial current
operator as
\begin{mathletters} \label{axial}
\begin{equation}
\delta X^{ia} = \frac{m_{bb'}^2}{32\pi^2 f^2}
\left( \ln \frac{m_{bb'}^2}{\lambda^2} + \frac{2}{3} \right)
[ X^{jb}, [ X^{ia}, X^{jb'} ] ]
- \frac{m_{bb'}^2}{32 \pi^2 f^2} \ln \frac{m_{bb'}^2}{\lambda^2}
{\cal O}^{i,bb'},
\end{equation}
by evaluating the loop integral using dimensional regularization with the
regularization scale $\lambda$ (see, e.g., Ref. \cite{BKM95}.), where
\begin{equation}
{\cal O}_\mu^{bb'} = g \{ [ {\textstyle\frac12} \lambda^b,
[ {\textstyle\frac12} \lambda^{b'}, {\textstyle\frac12} \lambda^a ] ]
\sigma_\mu \}
+ \frac{h}{N_c} \{ [ {\textstyle\frac12} \lambda^b, [
{\textstyle\frac12} \lambda^{b'}, {\textstyle\frac12} \lambda^a ] ] \}
\{ \sigma_\mu \},
\label{2ndax}
\end{equation}
which comes from Fig. \ref{fig:ax1}(b).
So the one-loop correction to the baryon axial current matrix elements
are obtained as
\end{mathletters}
\begin{equation}
\delta \alpha_{B'B}^a = \beta_{B'B}^{a,\Pi}
\frac{m_\Pi^2}{16\pi^2 f^2} \ln \frac{m_\Pi^2}{\lambda^2}
+ \tilde\beta_{B'B}^{a,\Pi} \frac{m_\Pi^2}{24\pi^2 f^2},
\label{beta}
\end{equation}
where $\Pi$ stands for $\pi$, $K$ and $\eta$ mesons.

The explicit results of $\beta_{B'B}^{a,\Pi}$ and
$\tilde \beta_{B'B}^{a,\Pi}$ with $g$ terms are given in Appendix B.
{}From these results, we can see that the one-loop corrections
are $O(1/N_c)$ at most since $f^2$ scales like $N_c$.
Furthermore, the corrections from Fig. \ref{fig:ax1}(b) are {\it of the
same order} as those of Fig. \ref{fig:ax1}(a).

\section{Baryon masses at tree level}

In this Section we investigate the baryon masses at tree level.
To estimate the baryon masses simultaneously in the $1/N_c$ expansion and
the chiral expansion, we must specify the relation between $1/N_c$ and
the pseudo-Goldstone boson mass $m_\Pi$.
In this paper, we use $m_\Pi \delta M \sim O(1)$ where $\delta M$ is the
octet-decuplet mass difference.
This gives $m_\Pi \sim O(\varepsilon)$ and $1/N_c \sim O(\varepsilon)$,
where $\varepsilon$ stands for a small expansion parameter.%
\footnote{This is consistent with the expansion of Ref. \cite{HHK97},
where the $\Delta$-nucleon mass difference is treated as small parameter
with the pion mass.}
A priori, there is no constraint on the relationship between $m_\Pi$ and
$N_c$.
In fact, the authors of Ref. \cite{BL96} used $m_\Pi^2 \delta M \sim O(1)$.
However as we shall see below, the leading order correction to the
degenerate baryon mass in the large $N_c$ limit is proportional to
$m_\Pi^2 N_c$ and we count it as $O(\varepsilon)$. This is consistent,
given that $m_\Pi\sim O(\varepsilon^2)$ in accordance with the
chiral expansion, and $N_c \sim O(\varepsilon^{-1})$.
We will compare our results with those of Ref. \cite{BL96} before
calculating the one-loop corrections.

The matrix elements of the effective Lagrangian which contribute to the
baryon mass can be written as
\begin{equation}
\langle {\cal L}_B \rangle =
 \sum_i \, (B| \tilde{\cal L}_{\rm eff}^{(i)} |B),
\end{equation}
where $\tilde{\cal L}_{\rm eff}^{(i)}$ represents that part of the
Lagrangian which can give a contribution of $O(\varepsilon^i)$.
Explicitly, these terms are
\begin{mathletters} \label{Lag:mass}
\begin{eqnarray}
\tilde{\cal L}^{(-1)}_{\rm eff} &=&
-a_0 \{ \bbox{1} \},
\label{lag-1} \\
\tilde{\cal L}^{(1)}_{\rm eff} &=&
- \frac{a_1}{N_c} \{ \sigma^j \} \{ \sigma^j \} - b_1 \{ m \},
\label{lag1} \\
\tilde{\cal L}^{(2)}_{\rm eff} &=&
- \frac{\alpha_1}{N_c} \,\mbox{tr}\, (m) \{ \bbox{1} \},
\label{lag2} \\
\tilde{\cal L}^{(3)}_{\rm eff} &=&
- \frac{b_2}{N_c} \{ m \sigma^j \} \{ \sigma^j \} - c_1 \{ m^2 \}
- \frac{c_2}{N_c} \{ m \} \{ m \},
\label{lag3} \\
\tilde{\cal L}^{(4)}_{\rm eff} &=&
- \frac{\alpha_2}{N_c^2} \,\mbox{tr}\, (m) \{ \sigma^j \} \{ \sigma^j \}
- \frac{\beta_1}{N_c} \,\mbox{tr}\, (m^2) \{ \bbox{1} \},
\label{lag4} \\
\tilde{\cal L}^{(5)}_{\rm eff} &=&
- \frac{c_3}{N_c} \{ m^2 \sigma^j \} \{ \sigma^j \}
- \frac{c_4}{N_c} \{ m \sigma^j \} \{ m \sigma^j \}
- \frac{c_5}{N_c} \{ m \} \{ m \sigma^j \} \{ \sigma^j \}
\nonumber \\ && \mbox{}
- d_1 \{ m^3 \} - \frac{d_2}{N_c} \{ m^2 \} \{ m \}
- \frac{d_3}{N_c^2} \{ m \} \{ m \} \{ m \},
\label{lag5}
\end{eqnarray}
up to $O(\varepsilon^5)$, where
\end{mathletters}
\begin{equation}
m = B_0 (\xi^\dagger m_q \xi^\dagger + \xi m_q \xi ).
\end{equation}
We make use of the standard relations between $B_0$ and squared pion and
kaon masses, $m_\pi^2 = 2B_0 \hat{m}$ and $m_K^2 = B_0 (\hat{m} + m_s)$,
where $\hat{m}$ is the average mass of $u$ and $d$ quarks and $m_s$ the
$s$-quark mass.
The quark mass matrix $m_q$ is given by
\begin{equation}
m_q = \hat{m}\, {\cal U} + m_s\, {\cal S},
\end{equation}
where
\begin{equation}
{\cal U} = \mbox{diag}(1,1,0), \qquad
{\cal S} = \mbox{diag}(0,0,1).
\end{equation}
Throughout this work, we assume SU(2) isospin symmetry with $m_u=m_d$.
Then, up to $O(\varepsilon^5)$, there are 15 low energy constants that
should be determined from experiments.
However, one can find that 6 terms give identical contributions to
all baryon masses so that 9 parameters remain which determine all baryon
mass differences.
In the following, we discuss the baryon masses at each order of $\varepsilon$.

{}From the Lagrangian (\ref{lag-1}), all octet and decuplet baryon masses
are degenerate at leading order, which gives the baryon mass
operator,
\begin{equation}
M_B^{(0)} = a_0 N_c,
\end{equation}
where the parameter $a_0$ sets the scale as a ``mass per color degree
of freedom''.

To the next order, the correction to the mass formula reads
\begin{equation}
\delta M_B^{(1)} = \frac{a_1}{N_c} \{ \sigma^j \} \{ \sigma^j \}
+ 2 B_0 \hat{m} N_c b_1.
\end{equation}
The $a_1$ term gives the splitting between octet and
decuplet while all states within the octet and decuplet are still
degenerate.
Although the original form of Eq. (\ref{lag1}) includes the operator
$\{ {\cal S} \}$, the resulting baryon masses do not depend on
strangeness since the expectation values of $\{ {\cal S} \}$ for our
baryon states are of $O(1)$ so that its contribution appears at the next
higher order.
Thus, at $O(\varepsilon^1)$, we get
\begin{eqnarray}
\delta M_8^{(1)} &=& \frac{3}{N_c} a_1 + 2 B_0 \hat{m} N_c b_1, \nonumber \\
\delta M_{10}^{(1)} &=& \frac{15}{N_c} a_1 + 2 B_0 \hat{m} N_c b_1,
\end{eqnarray}
where $M_8$ and $M_{10}$ denote the baryon octet and decuplet masses,
respectively.

At $O(\varepsilon^{2})$ there are two contributions.
One is from $\tilde{\cal L}^2_{\rm eff}$ of (\ref{lag2}) and
the other is from the remaining part of the $b_1$ term
of $\tilde{\cal L}^1_{\rm eff}$:
\begin{equation}
\delta M_B^{(2)} = 2 B_0 (m_s - \hat{m}) b_1 \{ {\cal S} \} +
2B_0 (m_s + 2\hat{m}) \alpha_1.
\end{equation}
It is clear that the $\alpha_1$ term gives the same mass shift to all baryons.
The dependence of the $b_1$ term on strangeness results from the SU(3)
flavor symmetry breaking and vanishes in the flavor SU(3) limit.
Therefore, up to this order, the baryon mass depends on total spin and
strangeness, but the $\Lambda$ and the $\Sigma$ are still degenerate.

The mass corrections at $O(\varepsilon^{3})$ can be obtained as
\begin{equation}
\delta M_B^{(3)} = \frac{2 B_0}{N_c} \hat{m} b_2 \{ \sigma^j \} \{ \sigma^j \}
+ \frac{2B_0}{N_c} (m_s - \hat{m}) b_2 \{ {\cal S} \sigma^j \} \{ \sigma^j \}
+ N_c (2B_0 \hat{m})^2 (c_1+c_2).
\end{equation}
The $b_2$ term involves two operators.
One of them depends on the total baryon spin and the other depends on the
spin of the strange quark(s).
As a result, the $\Sigma$ decouples from the $\Lambda$ at this order.
Up to this order, we have 4 operators in the baryon mass formula, namely,
$\{ {\bf 1} \}$, $\{ \sigma^j \} \{ \sigma^j \}$, $\{ {\cal S} \}$ and
$\{ {\cal S} \sigma^j \} \{ \sigma^j \}$.
The $c_1$ and $c_2$ terms give the same contributions to all baryons.
The matrix elements of the operators can be evaluated using
\begin{eqnarray}
\{ {\cal S} \} &=& (N_5+N_6), \nonumber \\
 \{ {\cal S} \} \{ {\cal S} \} &=& (N_5+N_6)^2,  \nonumber \\
\{ {\cal S} \sigma_j \} \{ \sigma_j \} &=& 2 (N_5+N_6) +
(N_5-N_6) [(N_1-N_2) + (N_3-N_4) + (N_5-N_6)]
+ 4 N_5 N_6 \nonumber \\
&& \mbox{} + 2 ( \alpha_1 \alpha^\dagger_2 \alpha^\dagger_5 \alpha_6
+ \alpha^\dagger_1 \alpha_2 \alpha_5 \alpha^\dagger_6
+ \alpha_3 \alpha^\dagger_4 \alpha^\dagger_5 \alpha_6 +
\alpha^\dagger_3 \alpha_4 \alpha_5 \alpha^\dagger_6 ),
\nonumber \\
\{ \sigma_j \} \{ \sigma_j \} &=& 2 [(N_1+N_2) + (N_3+N_4) + (N_5+N_6)]
+ 4(N_1 N_2 + N_3 N_4 + N_5 N_6)
\nonumber \\ && \mbox{}
+ [ (N_1-N_2) + (N_3-N_4) + (N_5-N_6)]^2
\nonumber \\
&& \mbox{} + 4( \alpha^\dagger_1 \alpha_2 \alpha_3 \alpha^\dagger_4
+ \alpha^\dagger_1 \alpha_2 \alpha_5 \alpha^\dagger_6
+ \alpha_1 \alpha^\dagger_2 \alpha^\dagger_3 \alpha_4
+ \alpha_1 \alpha^\dagger_2 \alpha^\dagger_5 \alpha_6
+ \alpha_3 \alpha^\dagger_4 \alpha^\dagger_5 \alpha_6
\nonumber \\ && \quad \mbox{}
+ \alpha_3 \alpha^\dagger_4 \alpha^\dagger_5 \alpha_6),
\nonumber \\
\{ {\cal S} \sigma_j \} \{ {\cal U} \sigma_j \} &=&
[(N_1-N_2) + (N_3-N_4)](N_5-N_6)
\nonumber \\ && \quad \mbox{}
+ 2 ( \alpha_1^\dagger \alpha_2 \alpha_5 \alpha_6^\dagger
+ \alpha_1 \alpha_2^\dagger \alpha_5^\dagger \alpha_6
+ \alpha_3^\dagger \alpha_4 \alpha_5 \alpha_6^\dagger
+ \alpha_3 \alpha_4^\dagger \alpha_5^\dagger \alpha_6).
\end{eqnarray}
All the matrix elements needed to calculate the baryon masses are given
in Table \ref{tab:me1}.

The explicit expression of mass corrections at $O(\varepsilon^{4})$ reads
\begin{eqnarray}
\delta M_B^{(4)} &=&
(2B_0)^2 [ (m_s^2 - \hat{m}^2) c_1
+ 2 \hat{m} (m_s - \hat{m}) c_2 ] \{ {\cal S} \}
\nonumber \\ && \mbox{}
+ \frac{\alpha_2}{N_c^2} 2B_0 (2 \hat{m} + m_s) \{ \sigma^j \} \{ \sigma^j \}
+ \beta_1 (2B_0)^2 ( 2 \hat{m}^2 + m_s^2 ).
\end{eqnarray}
The combination of $c_1$ and $c_2$ terms depends on the strangeness, and
the $\alpha_2$ term gives the next order contribution to the decuplet--octet
splitting.
Therefore, all of the above terms can be absorbed into the formulas valid
up to $O(\varepsilon^3)$.

Then, up to this order, we have three mass relations,
\begin{eqnarray}
&& (\mbox{\it M1}) \quad
M_\Delta - M_N = M_{\Sigma^*} - M_\Sigma + \frac32 ( M_\Sigma - M_\Lambda ),
\nonumber \\
&& (\mbox{\it M2}) \quad
3 M_\Lambda + M_{\Sigma} - 2 M_N - 2 M_\Xi  = 0,
\nonumber \\
&& (\mbox{\it M3}) \quad
M_{\Omega} - M_{\Xi^*} = M_{\Xi^*} - M_{\Sigma^*}
= M_{\Sigma^*} - M_\Delta,
\label{mass-rel1}
\end{eqnarray}
where ({\it M1}) is the hyperfine splitting rule, ({\it M2}) the
Gell-Mann--Okubo (GMO) relation and ({\it M3}) the decuplet equal
spacing (DES) rule.

The $O(\varepsilon^5)$ correction to the baryon mass has a more complicated
form:
\begin{eqnarray}
\delta M_B^{(5)} &=&
\frac{1}{N_c} (2B_0)^2 ( m_s - \hat{m} )^2 c_2 \{ {\cal S} \} \{ {\cal S} \}
\nonumber \\ && \mbox{}
+ \frac{c_3}{N_c} (2B_0)^2 \left[ \hat{m}^2 \{ \sigma^j \} \{ \sigma^j \}
+ (m_s^2 - \hat{m}^2) \{{\cal S} \sigma^j \} \{ \sigma^j \} \right]
\nonumber \\ && \mbox{}
+ \frac{c_4}{N_c} (2B_0)^2 \left[ \hat{m}^2 \{ \sigma^j \} \{ \sigma^j \}
+ (m_s^2 - \hat{m}^2) \{ {\cal S} \sigma^j \} \{ \sigma^j \}
- (m_s - \hat{m})^2 \{ {\cal U} \sigma^j \} \{ {\cal S} \sigma^j \} \right]
\nonumber \\ && \mbox{}
+ \frac{c_5}{N_c} (2B_0)^2 \left[ \hat{m}^2
+ \hat{m} (m_s - \hat{m}) \{ {\cal S} \sigma^j \} \{ \sigma^j \} \right]
\nonumber \\ && \mbox{}
+ N_c (d_1+d_2+d_3) (2B_0 \hat{m})^3.
\label{Tree-5}
\end{eqnarray}
There are more terms including $c_5$ and $d_{1,2,3}$, but they give
contributions only at higher orders.
The mass formula (\ref{Tree-5}) includes the operators
$\{ {\cal S} \} \{ {\cal S} \}$
and $\{ {\cal U} \sigma^j \} \{ {\cal S} \sigma^j \}$ in addition to the
operators that appeared already at the lower order.
Because of these new operators, the mass relations ({\it M2}) and
({\it M3}) of Eq. (\ref{mass-rel1}) are modified, whereas ({\it M1})
is still valid.
Instead of ({\it M2}) and ({\it M3}), we find improved GMO and DES
rules \cite{DJM94,Luty95}:
\begin{eqnarray}
&& (\mbox{\it M2}') \quad
3M_\Lambda + M_\Sigma - 2( M_N + M_\Xi) =
(M_{\Sigma^*} - M_\Delta) - (M_\Omega - M_{\Xi^*}),
\nonumber \\
&& (\mbox{\it M3}') \quad
(M_\Omega - M_{\Xi^*}) - (M_{\Xi^*} - M_{\Sigma^*})
= (M_{\Xi^*} - M_{\Sigma^*}) - (M_{\Sigma^*} - M_\Delta),
\end{eqnarray}
which work better than ({\it M2}) and ({\it M3}). Empirically,
the left and right hand sides of ({\it M1}) give $(293=308)$, and
$({\it M2})$ and $({\it M3})$, respectively,
lead to $(27 = 0)$ and $(139=149=152)$, whereas $({\it M2}')$ gives $(27=11)$
and $({\it M3}')$ gives $(-3=-8)$, where the numbers are given in MeV.
Combining these relations with
$(\it M1)$ gives 
\begin{eqnarray}
M_{\Xi^*} - M_\Xi &=& M_{\Sigma^*} - M_\Sigma,
\label{addmr} \\
(215 &=& 192) \nonumber
\end{eqnarray}
where the numbers show again the experimental values.
Note that this is not an independent mass relation.
The modified DES rule $(\it M3')$ was first derived by Okubo
\cite{Okub63a} in the form of
\begin{equation}
M_\Omega - M_\Delta = 3(M_{\Xi^*} - M_{\Sigma^*}),
\end{equation}
which is just a re-combination of $(\it M1)$, $(\it M2')$ and $(\it M3')$.

\begin{figure}
\centerline{\epsfig{file=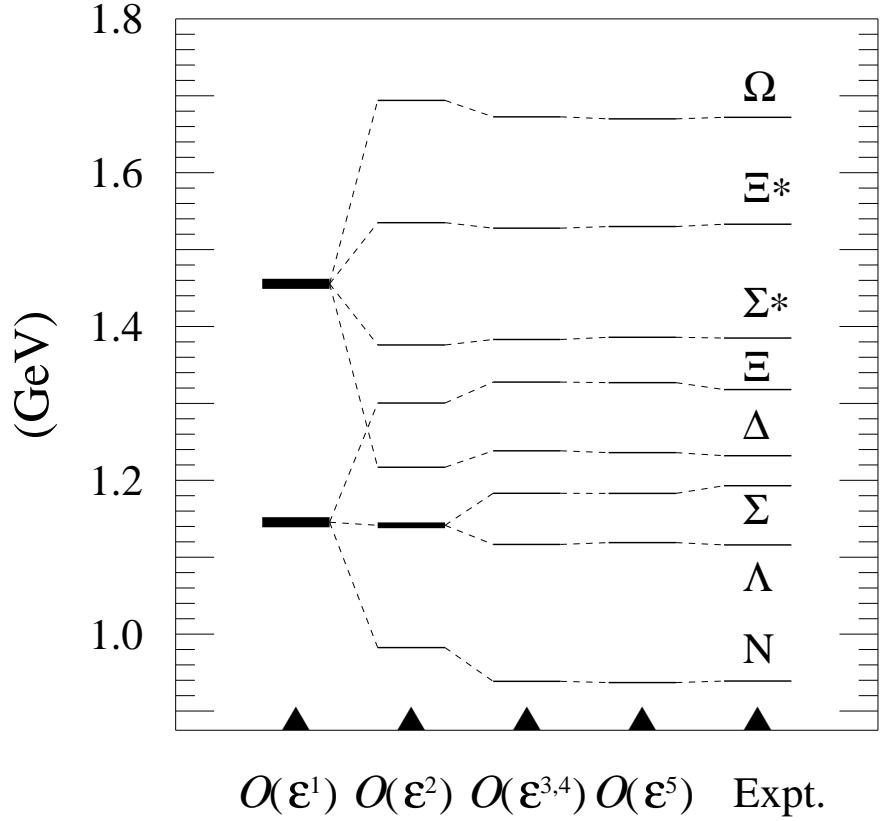, width=0.7\hsize}}
\caption{Best fit of baryon masses (tree) up to $O(\varepsilon^5)$.
Thick lines represent degenerate states.}
\label{fig:sp}
\end{figure}

Since there are 6 different types of operators up to $O(\varepsilon^5)$,
we can write the mass formula in a compact form as
\begin{equation}
M_B = a + b \{ \sigma^j \} \{ \sigma^j \} + c \{ {\cal S} \}
+ d \{ {\cal S} \sigma^j \} \{ \sigma^j \} + e \{ {\cal S} \} \{ {\cal S} \}
+ f \{ {\cal U} \sigma^j \} \{ {\cal S} \sigma^j \},
\label{massform}
\end{equation}
where the $c$ term comes in at $O(\varepsilon^2)$, the $d$ term at
$O(\varepsilon^3)$, and the $e$ and $f$ terms at $O(\varepsilon^5)$.
The best $\chi^2$ fits to the baryon masses up to $O(\varepsilon^5)$ are
shown in Table \ref{tab:fit} and Fig. \ref{fig:sp}.
The best fit up to $O(\varepsilon^4)$ is the same as that of $O(\varepsilon^3)$.
This is because the mass formula of $O(\varepsilon^4)$ does not introduce
any new operator. A reasonable baryon mass
spectrum is already found at $O(\varepsilon^3)$. Corrections from the
$O(\varepsilon^5)$ operators are evidently not so important.
Note also that the coefficients of the operators involving ${\cal S}$
include a factor $(m_s - m)$ so that the $c$, $d$, $e$ and $f$ terms of
Eq. (\ref{massform}) vanish in the limit of exact SU(3) flavor symmetry.

Before proceeding to the loop corrections, let us compare our results with
those of Ref. \cite{BL96}, which uses different counting so that
$O(\epsilon') = O(m_q) = O(1/N_c)$.
At the leading order, the authors of Ref. \cite{BL96} obtained 5 mass
relations:
\begin{eqnarray}
&& (M_\Xi - M_\Sigma) - (M_\Sigma - M_N) +
\frac32 (M_\Sigma - M_\Lambda) = 0 \quad (= -13.5),
\nonumber \\
&& (M_{\Xi^*} - M_{\Sigma^*}) - (M_{\Sigma^*} - M_\Delta) = 0 \quad (= -8),
\nonumber \\
&& (M_\Omega - M_{\Xi^*}) - (M_{\Xi^*} - M_{\Sigma^*}) = 0 \quad (=-3),
\nonumber \\
&& (M_\Sigma - M_N) - (M_\Lambda - M_N) = 0 \quad (=77), \nonumber \\
&& (M_{\Sigma^*} - M_\Delta) - (M_\Lambda - M_N) = 0 \quad (=-24),
\end{eqnarray}
where the numbers in parenthesis on the right hand side are the empirical
ones in MeV.
The first three relations are re-combinations of $(\it M1)$, $(\it M2)$
and $(\it M3)$ and they are reasonably consistent with experiments. However,
the deviations of the last two relations are larger compared with the
first three relations.
In our scheme, this discrepancy can be understood easily because the first
three relations hold up to $O(\varepsilon^3)$ and $O(\varepsilon^4)$
whereas the last two hold only up to $O(\varepsilon^2)$.

\section{Loop Corrections to Baryon masses}

\begin{figure}
\centering
\epsfig{file=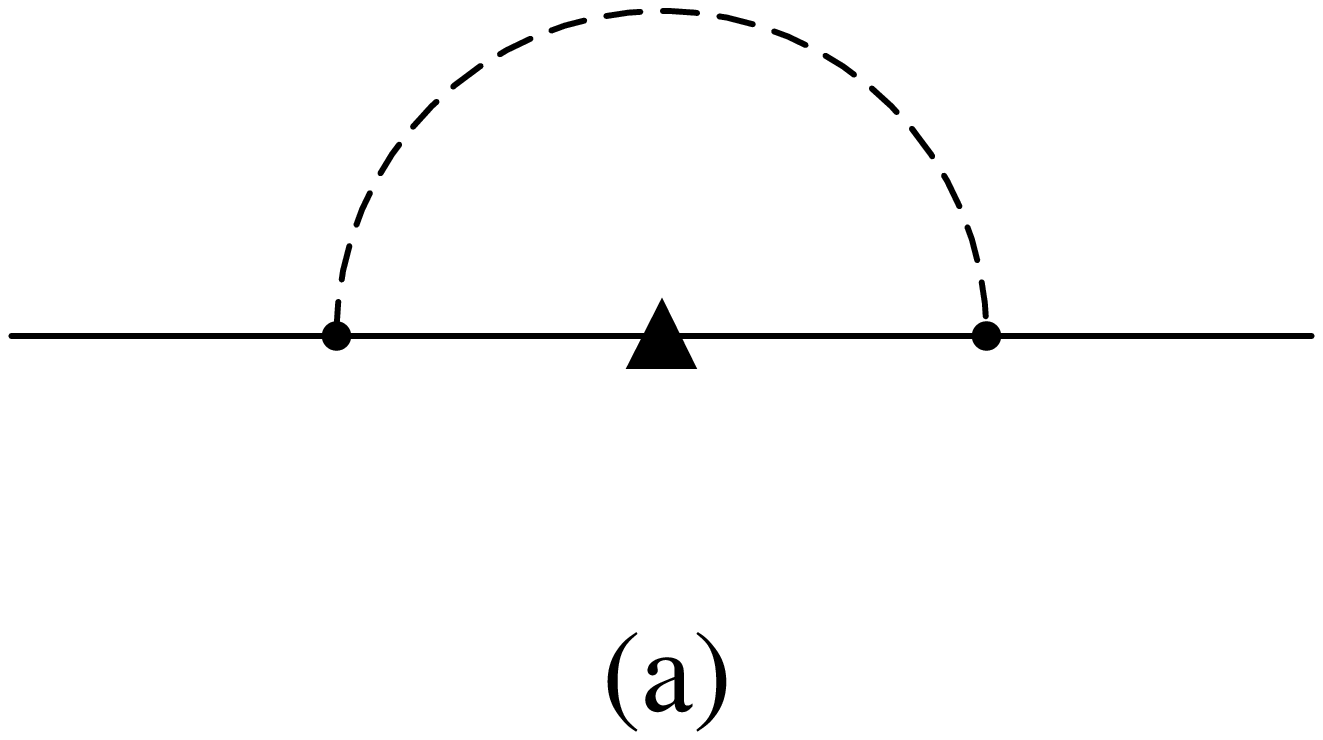, width=0.4\hsize} \qquad
\epsfig{file=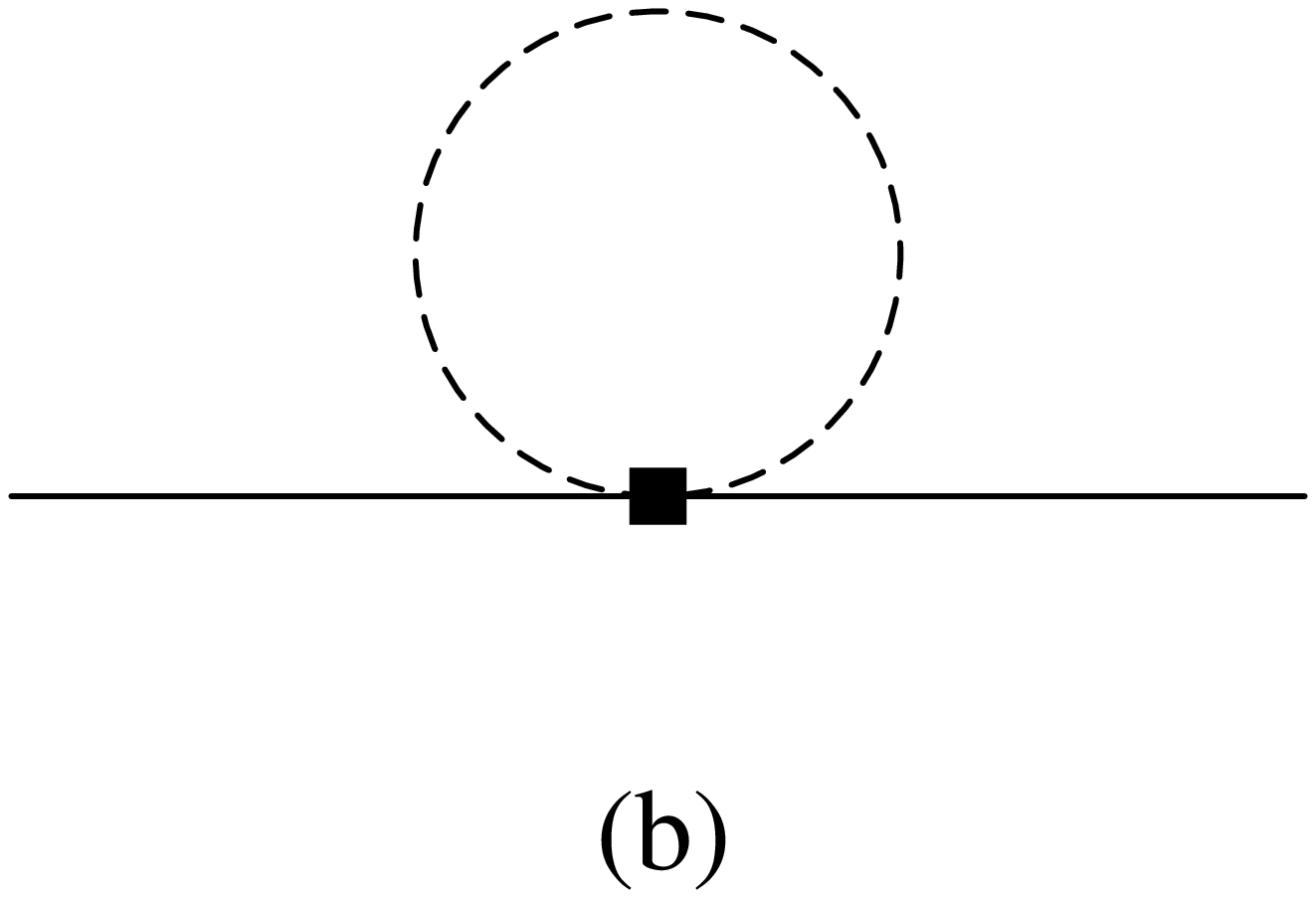, width=0.4\hsize}
\caption{One-loop corrections to the baryon mass. The filled-triangle
denotes the mass insertion to the intermediate baryon state and the
filled-box represents the meson-meson-baryon-baryon coupling from the
chiral Lagrangian of (\protect\ref{Lag:mass}) and (\protect\ref{Lag:high}).}
\label{fig:ml}
\end{figure}

The one-loop corrections to the baryon masses are obtained from the
diagrams shown in Fig. \ref{fig:ml}. First, let us consider the diagram
of Fig. \ref{fig:ml}(a) without mass insertions to the intermediate
baryon states, which corresponds to Fig. \ref{fig:wf1}(a).
At first glance, this one-loop correction appears to be inconsistent with the
$1/N_c$ expansion.
Since each vertex carries a factor $\sqrt{N_c}$, the one-loop correction
is $O(N_c)$.
A similar feature occurs in the case of the baryon axial current, where
the wave function renormalization part must be included to give
the proper commutator structure which is essential to be consistent with
the $1/N_c$ expansion.
In the case of the baryon self energy, however, there is no other term that
can lead to this commutator structure.
Thus the one-loop correction {\it is not suppressed\/} as compared to the
tree level baryon masses \cite{LM94}.
In fact, this one-loop correction starts from $O(N_c)$, but we can see
that the corrections of this order are the same to all baryons so that
it can be absorbed in the $a_0$ term of the baryon mass.

The one-loop baryon self energy is obtained as
\begin{equation}
\Sigma_B (\omega) = \frac{i}{2f^2} (B| X^{\mu a} X^{\nu a} |B)
\int \frac{d^4 k}{(2\pi)^4} \frac{i}{(p-k)\cdot v}
\frac{i}{k^2 - m_{aa}^2} ( -k^\mu k^\nu ),
\end{equation}
where $\omega = p \cdot v$ and $X^{\mu a}$ is defined in Eq. (\ref{Xia}).
After evaluating the loop integral we find:
\begin{mathletters} \label{loop0}
\begin{equation}
\delta M_B = - \frac{m_\Pi^3}{16\pi f^2} \langle {\cal O}^\Pi (B) \rangle,
\end{equation}
with
\begin{equation}
\langle {\cal O}^\Pi (B) \rangle = (B| {\cal O}^\Pi |B) =
\frac23 (B| \sum_a X^{ia} X^{ia} |B),
\label{loop0-2}
\end{equation}
where $a=1,2,3$ for the pion loop, $a=4,\dots,7$ for the kaon loop,
and $a=8$ for the eta loop.
The operator ${\cal O}^\Pi$ can be computed straightforwardly to give
\end{mathletters}
\begin{mathletters} \label{Op-O}
\begin{eqnarray}
{\cal O}^\pi &=& g^2 \left[
 \frac{N_c^2}{2} + 2N_c - \left( N_c + \frac83 \right) \{ {\cal S} \}
+ \frac16 \{ {\cal S} \} \{ {\cal S} \} - \frac13 \{ \sigma^j \} \{ \sigma^j \}
+ \frac23 \{ {\cal S} \sigma^j \} \{ \sigma^j \} \right]
\nonumber \\ && \mbox{} +
 \frac{gh}{3N_c} \left[ N_c - \{ {\cal S} \} + 2 \right]
\left[ \{ \sigma^j \} \{ \sigma^j \} - \{ {\cal S} \sigma^j \} \{ \sigma^j \}
\right]
\nonumber \\ && \mbox{} +
\frac{h^2}{6N_c^2} \left[ \{ \sigma^i \} \{ \sigma^i \}
- 2 \{ {\cal S} \sigma^i \} \{ \sigma^i \} + 2 \{ {\cal S} \}
+ \{ {\cal S} \} \{ {\cal S} \} \right] \{ \sigma^j \} \{ \sigma^j \},
\\
{\cal O}^K &=& g^2 \left[
N_c + \left( N_c + \frac53 \right) \{ {\cal S} \}
- \frac13 \{ {\cal S} \sigma^j \} \{ \sigma^j \}
- \frac23 \{ {\cal S} \} \{ {\cal S} \} \right]
\nonumber \\ && \mbox{} +
\frac{2gh}{3N_c} \left[
( \{ {\cal S} \} + 1 ) \{ \sigma^j \} \{ \sigma^j \}
+ ( N_c - 2\{ {\cal S} \} + 1 ) \{ {\cal S} \sigma^j \} \{ \sigma^j \}
\right]
\nonumber \\ && \mbox{} +
\frac{h^2}{3N_c^2} \left[ N_c + (N_c-1) \{ {\cal S} \}
+ \{ {\cal S} \sigma^i \} \{ \sigma^i \} - 2 \{ {\cal S} \} \{ {\cal S} \}
\right] \{ \sigma^j \} \{ \sigma^j \},
\\
{\cal O}^\eta &=& g^2 \left[
\{ {\cal S} \} + \frac12 \{ {\cal S} \} \{ {\cal S} \}
- \frac13 \{ {\cal S} \sigma^j \} \{ \sigma^j \}
+ \frac{1}{18} \{ \sigma^j \} \{ \sigma^j \} \right]
\nonumber \\ && \mbox{} +
\frac{gh}{9N_c} [ N_c -  3 \{ {\cal S} \} ]
\left[ \{ \sigma^j \} \{ \sigma^j \}
- 3 \{ {\cal S} \sigma^j \} \{ \sigma^j \} \right]
\nonumber \\ && \mbox{} +
\frac{h^2}{18N_c^2} \left[ N_c - 3 \{ {\cal S} \} \right]^2
\{ \sigma^j \} \{ \sigma^j \}.
\end{eqnarray}
\end{mathletters}

There are several remarks concerning this result.
As we discussed before, the pion loop correction ${\cal O}^\pi$
includes the factor $N_c^2$, which gives a correction of order $N_c$
when combined with the factor $1/f^2$.
Thus it is not consistent with the $1/N_c$ expansion.
However this term has a trivial operator structure and therefore does not
contribute to the baryon mass differences.
Furthermore, because of $m_\Pi^3$, it is of $O(\epsilon^2)$ and suppressed in
comparison with the leading order of the tree level mass.
Secondly, the leading orders of ${\cal O}^\pi$, ${\cal O}^K$ and
${\cal O}^\eta$ are, respectively, $O(N_c^2)$, $O(N_c^1)$ and $O(N_c^0)$.
The leading order in $1/N_c$ of each term is given in Table \ref{taborder}.
One would expect that the $gh$ and $h^2$ terms are suppressed as
compared to the $g^2$ term.
This is true for the pion and kaon loop corrections as can
be seen from Table \ref{taborder}.
However, in the case of $\eta$-meson loop, the $gh$ and $h^2$
terms are as the same order as the $g^2$ term.
This is similar to what we have seen in the $\alpha^8$ calculation in
Section III.
Thus in order to get the correct result for the $\eta$ loop corrections,
we have to consider $n$-body operators in general, unless the coupling
constants of such operators are numerically suppressed.
In our estimate, we keep terms up to the 2-body operator in
$X^{ia}$, i.e., the $h$ term.
Finally, we note that the $g^2$ terms involve the operators, $\{ {\cal S} \}$,
$\{ {\cal S} \}\{ {\cal S} \}$, $\{ \sigma^j \} \{ \sigma^j \}$ and
$\{ {\cal S} \sigma^j \} \{ \sigma^j \}$, which have already appeared in the
mass formula (\ref{massform}).
This means that the $g^2$ terms satisfy the three mass relations,
$(\it M1)$, $(\it M2')$, and $(\it M3')$.%
\footnote{Note however that the mass relations $(\it M2)$ and $(\it M3)$
receive corrections from the $g^2$ term.}
Corrections to the mass relations come from the $gh$ and $h^2$ terms
which include $\{ {\cal S} \} \{ \sigma^j \} \{ \sigma^j \}$, etc.

To estimate the loop correction from Fig. \ref{fig:ml}(a), we
include the mass insertions to the intermediate baryon states.
Let the mass difference be denoted by $\delta M_{B'}$.
Then the baryon self energy from this diagram reads
\begin{equation}
\Sigma(\omega) = - \frac{1}{f^2} (B| X^{\mu a} |B') (B'| X^{\nu a} |B)
\tilde{\cal I}_{\mu\nu} (\omega),
\end{equation}
where
\begin{equation}
\tilde{\cal I}_{\mu\nu} (\omega) = -i \int \frac{d^4 k}{(2\pi)^4}
\left( \frac{1}{k \cdot v - \omega + \delta M_{B'}} \right)
\frac{k_\mu k_\nu}{m_{aa}^2 - k^2}.
\end{equation}
Calculation of the loop integral gives
\begin{mathletters}
\label{loop1}
\begin{equation}
\delta M_B = {\cal J}_2 (\delta M_{B'}, m_\Pi) \, \gamma_{B'}^\Pi (B) ,
\end{equation}
where
\begin{equation}
\gamma_{B'}^\Pi (B) = \sum_a ( B| X^{ia} | B') (B'| X^{ia} |B),
\end{equation}
with $a=1,2,3$ for the pion loop, $a=4,\dots,7$ for the kaon loop and
$a=8$ for the eta loop, and
\begin{eqnarray}
{\cal J}_2 (x,m_A) &=& - \frac{x m_A^2}{48\pi^2 f^2}
\left\{ 2 - 3\ln \left( \frac{m_A}{\lambda} \right)^2 \right\}
 - \frac{1}{12\pi^2 f^2} (m_A^2 - x^2)^{3/2} \arccos \frac{x}{m_A}
\nonumber \\ && \mbox{}
+ \frac{x^3}{24\pi^2 f^2}
\left\{ 1 - \ln \left( \frac{m_A}{\lambda} \right)^2 \right\},
\qquad \mbox{ for $m_A^2 > x^2$},
\nonumber \\
&=& - \frac{x m_A^2}{48\pi^2 f^2}
\left\{ 2 - 3\ln \left( \frac{m_A}{\lambda} \right)^2 \right\}
 + \frac{1}{24\pi^2 f^2} (x^2 - m_A^2)^{3/2}
\ln \frac{x - \sqrt{x^2-m_A^2}}{x + \sqrt{x^2-m_A^2}}
\nonumber \\ && \mbox{}
+ \frac{x^3}{24\pi^2 f^2}
\left\{ 1 - \ln \left( \frac{m_A}{\lambda} \right)^2 \right\},
\qquad \mbox{ for $m_A^2 < x^2$}.
\label{int-J}
\end{eqnarray}
In the limit $\delta M_{B'}=0$, we can recover the result (\ref{loop0}).
In the case of $\delta M_{B'}=0$ (or constant), the loop correction can be
represented in terms of the operators given in Eq. (\ref{loop0-2}).
This is possible because the loop integral does not depend on the
intermediate baryon state.
However, this is not the case in Eq. (\ref{loop1}) since the loop
integral depends on $\delta M_{B'}$.

\end{mathletters}

We can write Eq. (\ref{loop1}) in a more convenient form as follows.
With the usual definitions,
\begin{equation}
\sigma^{\pm 1} = \mp \frac{1}{\sqrt{2}} \left( \sigma^x \pm i \sigma^y
\right), \qquad 
\sigma^0 = \sigma^z,
\end{equation}
we use the Wigner--Eckart theorem,
\begin{equation}
(\gamma',\, j',\, m'\, | X(k,q) | \gamma,\, j,\, m)
= (-1)^{2k} \frac{ (j,m,k,q| j',m') }{ \sqrt{2j'+1} }
(\gamma' \, j' || X(k) || \gamma\, j).
\end{equation}
Then after some algebra, one can show that
\begin{eqnarray}
\gamma_{B'}^\pi (B) &=& \frac{c_B}{2} \sum_{a=1\pm i2, 3}
( 1 + \delta^{a3} ) \,[ (B' || X^a || B) ]^2,
\nonumber \\
\gamma_{B'}^K (B) &=& \frac{c_B}{2} \sum_{a=4\pm i5, 6 \pm i7}
\,[ (B' || X^a || B) ]^2,
\nonumber \\
\gamma_{B'}^\eta (B) &=& c_B \,[ (B' || X^8 || B) ]^2,
\end{eqnarray}
where $c_B = 1/2$ for octet baryons and $1/4$ for decuplet baryons.
Since
\begin{equation}
X^{1+i2,1+i2} = -g \sqrt{2} \alpha_1^\dagger \alpha_4
- \frac{h}{N_c} \sqrt{2}
(\alpha_1^\dagger \alpha_3 + \alpha_2^\dagger \alpha_4)
(\alpha_1^\dagger \alpha_2 + \alpha_3^\dagger \alpha_4 + \alpha_5^\dagger
\alpha_6),
\end{equation}
and so on, one can compute the matrix elements $\gamma_{B'}^\Pi (B)$ using
the baryon wave functions given in Appendix A.
The final results for $\gamma_{B'}^\Pi (B)$ are given in Appendix C.

By comparison with Eq. (\ref{loop0}), we therefore have the relation
\begin{equation}
{\cal O}^\Pi (B) = \frac{2}{3} \sum_{B'} \gamma_{B'}^\Pi (B),
\label{closure}
\end{equation}
which can be obtained by taking $\delta M_{B'}=0$ in Eq. (\ref{loop1}).
However, by inserting $\gamma_{B'}^\Pi (B)$ given in Appendix C, one can
find that the above closure relation does {\em not} hold with
$B' \in \{ \underline{\bf 8} \}$ and $\{ \underline{\bf 10} \}$ only.
This is because we have
\begin{equation}
1 \neq \sum_{B'=\{ \underline{\bf 8} \}, \{ \underline{\bf 10} \}}
| B') (B'|,
\end{equation}
in the large $N_c$ limit.
The equality in the closure relation holds only for $N_c=3$.
To form a complete set, we need an infinite number of states for infinite
$N_c$.
However, fortunately in our case, $X^{ia}$ is a spin-1 operator.
So what we need in order to satisfy the relation (\ref{closure}) is to
include the intermediate baryon states up to spin 5/2.
This is done in Appendix A, where we give all the states $B'$
of spin 1/2, 3/2, and 5/2 to fulfill Eq. (\ref{closure}).
All these additional states are fictitious, i.e., they do not exist in the
real world with $N_c=3$, but they are needed to satisfy the closure
relation in the large $N_c$ limit.
Note also that the baryon self-energy of Eq. (\ref{loop1}) starts at
$O(\varepsilon^2)$.

The contribution to the baryon self energy from Fig. \ref{fig:ml}(b)
vanishes for the meson-baryon couplings (\ref{Lag:mb}).
The contribution of such a diagram comes from the effective Lagrangian
(\ref{Lag:mass}).
Consider for example the one-loop correction from
$\tilde{\cal L}^{(1)}_{\rm eff}$ of Eq. (\ref{lag1}) to the baryon self
energy.
This one-loop correction comes from the $\{ m \}$ term of $\tilde{\cal
L}^{(1)}_{\rm eff}$, which is expanded as
\begin{equation}
m = m_q - \frac{1}{2f^2} [ \Pi, [ \Pi, m_q ]_+ ]_+ + \dots,
\end{equation}
where $ [ A, B ]^{}_+ = AB + BA $.
Then the one-loop correction to the baryon self-energy reads
\begin{equation}
\Sigma(\omega) = - \frac{b_1}{2f^2} \{ [ \Pi, [ \Pi, m_q ]_+ ]_+ \}
\Delta_\Pi,
\end{equation}
where
\begin{equation}
\Delta_\Pi = -i \int \frac{d^4 k}{(2\pi)^4} \frac{1}{m_\Pi^2 - k^2}.
\end{equation}
By evaluating the loop integral using dimensional regularization, we get
\begin{equation}
\delta M_B = \frac{m_\Pi^2}{16 \pi^2 f^2} \ln \frac{m_\Pi^2}{\lambda^2}
(B| {\cal P}^\Pi_1 |B),
\end{equation}
where
\begin{equation}
(B| {\cal P}^\Pi_1 |B) = - \frac{b_1}{2} \sum_a
\{ [ {\textstyle\frac12} \lambda^a , [ {\textstyle\frac12}
\lambda^a, m_q ]^{}_+ ]^{}_+ \}.
\end{equation}
In the same way, we can compute the baryon self energy of
Fig. \ref{fig:ml}(b) from the higher order terms of Eq. (\ref{Lag:mass})
to obtain
\begin{equation}
\delta M_B = \frac{m_\Pi^2}{16 \pi^2 f^2} \ln \frac{m_\Pi^2}{\lambda^2}
(B| {\cal P}^\Pi |B),
\end{equation}
where
\begin{eqnarray}
{\cal P}^\Pi &=& \sum_a \biggl[
- \frac{b_1}{2} \{ [ {\textstyle\frac12} \lambda^a,
[ {\textstyle\frac12} \lambda^a, m_q ]_+ ]_+ \}
- \frac{\alpha_1}{2} \,\mbox{tr}\, \left( [ {\textstyle\frac12} \lambda^a,
[ {\textstyle\frac12} \lambda^a, m_q ]_+ ]_+ \right)
\nonumber \\ && \mbox{}
- \frac{b_2}{8N_c}
\{ [ \lambda^a , [ \lambda^a , m_q ]^{}_+ ]^{}_+ \sigma^i \} \{ \sigma^i \}
- \frac{c_1}{4} \{ m_q [ \lambda^a , [ \lambda^a , m_q ]^{}_+ ]^{}_+ \}
\nonumber \\ && \mbox{}
 - \frac{c_2}{4N_c} \{ m_q \}
\{ [ \lambda^a , [ \lambda^a , m_q ]^{}_+ ]^{}_+ \},
- \frac{\alpha_2}{8N_c^2} \,\mbox{tr}\, \left( [ \lambda^a ,
[ \lambda^a , m_q ]^{}_+ ]^{}_+ \right) \{ \sigma^j \} \{ \sigma^j \}
\nonumber \\ && \mbox{}
- \frac{\beta_1}{4} \,\mbox{tr}\, \left( m_q [ \lambda^a ,
[ \lambda^a , m_q ]^{}_+ ]^{}_+ \right) \biggr].
\end{eqnarray}
Explicit calculation gives
\begin{mathletters} \label{Op-P}
\begin{eqnarray}
{\cal P}^\pi &=& - \frac32 b_1 (2B_0 \hat{m}) [ N_c - \{ {\cal S} \} ]
 - 3 (2B_0 \hat{m}) \alpha_1
\nonumber \\ && \mbox{}
 - \frac{3}{2N_c} (2B_0 \hat{m}) \left[ \{ \sigma^i \} \{ \sigma^i \}
              - \{ {\cal S} \sigma^i \} \{ \sigma^i \} \right] b_2
 - 3 (2B_0 \hat{m})^2 \left[ N_c - \{ {\cal S} \} \right] c_1
\nonumber \\ && \mbox{}
 - \frac{3}{N_c} (2B_0 \hat{m}) (2B_0)
     \left[ \hat{m} N_c + (m_s - \hat{m}) \{ {\cal S} \} \right]
     \left[ N_c - \{ {\cal S} \} \right] c_2
\nonumber \\ && \mbox{}
 - \frac{3}{N_c^2} (2B_0) \hat{m} \{ \sigma^j \} \{ \sigma^j \} \alpha_2
 - 6 (2B_0 \hat{m})^2 \beta_1,
\\
{\cal P}^K &=& - \frac12 b_1 (2B_0) (\hat{m}+m_s) [ N_c + \{ {\cal S} \} ]
 - 2 (2B_0) ( \hat{m} + m_s ) \alpha_1
\nonumber \\ && \mbox{}
 - \frac{1}{2N_c} (2B_0) (\hat{m} + m_s) \left[ \{ \sigma^i \} \{ \sigma^i \}
             + \{ {\cal S} \sigma^i \} \{ \sigma^i \} \right] b_2
\nonumber \\ && \mbox{}
 - (2B_0) (\hat{m} + m_s) (2B_0)
        \left[ \hat{m} N_c + (2m_s - \hat{m}) \{ {\cal S} \} \right] c_1
\nonumber \\ && \mbox{}
 - \frac{1}{N_c} (2B_0) (\hat{m} + m_s)
          \left[ \hat{m} N_c + (m_s - \hat{m}) \{ {\cal S} \} \right]
          \left[ N_c + \{ {\cal S} \} \right] c_2
\nonumber \\ && \mbox{}
 - \frac{2}{N_c^2} (2B_0) (m_s + \hat{m}) \{ \sigma^j \} \{ \sigma^j \} \alpha_2
 - 2(2B_0)^2 (m_s + \hat{m})^2 \beta_1,
\\
{\cal P}^\eta &=& - \frac16 b_1 (2B_0)
              [ \hat{m} N_c + (4m_s - \hat{m}) \{ {\cal S} \} ]
 - \frac13 (2B_0) ( \hat{m} + 2 m_s ) \alpha_1
\nonumber \\ && \mbox{}
 - \frac{1}{6N_c} (2B_0) \left[ \hat{m} \{ \sigma^i \} \{ \sigma^i \}
      + (4m_s - \hat{m}) \{ {\cal S} \sigma^i \} \{ \sigma^i \} \right] b_2
\nonumber \\ && \mbox{}
 - \frac13 (2B_0)^2 \left[ \hat{m}^2 N_c + (4m_s^2 - \hat{m}^2)
           \{ {\cal S} \} \right] c_1
\nonumber \\ && \mbox{}
 - \frac{1}{3N_c} (2B_0)^2
       \left[ \hat{m} N_c + (m_s - \hat{m}) \{ {\cal S} \} \right]
       \left[ \hat{m} N_c + (4m_s - \hat{m}) \{ {\cal S} \} \right] c_2
\nonumber \\ && \mbox{}
 - \frac{1}{3N_c^2} (2B_0) (2m_s + \hat{m} ) \{\sigma^j\} \{\sigma^j\} \alpha_2
 - \frac23 (2B_0)^2 ( 2m_s^2 + \hat{m}^2 ) \beta_1.
\end{eqnarray}
Thus the leading order of this loop correction is $O(\varepsilon^4)$.
\end{mathletters}

However, there can be other one-loop corrections at $O(\varepsilon^4)$ from
higher order terms in the chiral Lagrangian, which can be written as
\begin{equation}
\delta {\cal L}_{\rm eff} = A_1 \{ A^\mu A_\mu \}
+ \frac{A_2}{N_c} \{ A_\mu \} \{ A^\mu \}.
\label{Lag:high}
\end{equation}
Generally, terms which involve $\{ (v \cdot A)^2 \}$
and $\{ v \cdot A \} \{ v \cdot A \}$ are also possible.
However, these terms can be absorbed
into Eq. (\ref{Lag:high}) because of the following identity in dimensional
regularization \cite{BM97}:
\begin{equation}
\int \frac{d^d k}{(2\pi)^d} \frac{(v \cdot k)^2}{m^2 - k^2}
= \frac{1}{d} \int \frac{d^d k}{(2\pi)^d} \frac{k^2}{m^2 - k^2}.
\end{equation}
The Lagrangian (\ref{Lag:high}) gives the one-loop correction of the type
of Fig. \ref{fig:ml}(b) as
\begin{equation}
\delta M_B = 
- \frac{m_\Pi^4}{16\pi^2 f^2} \ln \frac{m_\Pi^2}{\lambda^2}
(B| {\cal Q}^\Pi |B),
\end{equation}
where
\begin{equation}
{\cal Q}^\Pi = \sum_a \biggl[ \frac{A_1}{4} \{ \lambda^a \lambda^a \}
+ \frac{A_2}{4N_c} \{ \lambda^a \} \{ \lambda^a \} \biggr].
\end{equation}
The leading order of this term is $O(\varepsilon^4)$ since
\begin{mathletters} \label{Op-Q}
\begin{eqnarray}
{\cal Q}^\pi &=& \frac{3 A_1}{4} ( N_c - \{ {\cal S} \} )
+ \frac{A_2}{4N_c} \left[ \{ \sigma^i \} \{ \sigma^i \}
- 2 \{ {\cal S} \sigma^i \} \{ \sigma^i \} + 2 \{ {\cal S} \}
+ \{ {\cal S} \} \{ {\cal S} \} \right],
\\
{\cal Q}^K &=& \frac{A_1}{2} ( N_c + \{ {\cal S} \} )
+ \frac{A_2}{2N_c} \left[ N_c + (N_c-1) \{ {\cal S} \}
+ \{ {\cal S} \sigma^i \} \{ \sigma^i \} - 2 \{ {\cal S} \} \{ {\cal S} \}
\right],
\\
{\cal Q}^\eta &=& \frac{A_1}{12} ( N_c + 3 \{ {\cal S} \} )
+ \frac{A_2}{12N_c} \left[ N_c - 3 \{ {\cal S} \} \right]^2.
\end{eqnarray}
\end{mathletters}

{}From the expressions for the operators ${\cal P}^\Pi$ and ${\cal Q}^\Pi$ of
Eqs.  (\ref{Op-P}) and (\ref{Op-Q}), we can see that these are linear
combinations of the operators that appeared already in Eq. (\ref{massform}).
This means that the loop corrections of Fig. \ref{fig:ml}(b) satisfy the mass
relations, $(\it M1)$, $(\it M2')$ and $(\it M3')$.

In addition to the one-loop corrections of the previous calculation,
we have to consider one more contribution, i.e., the $1/M_B$ corrections.
They have been calculated in Refs. \cite{BM96a,BM97} within the framework
of baryon chiral perturbation theory.
To estimate the $1/M_B$ corrections, one can use the relativistic form
of the effective Lagrangian and then expand it to obtain the $1/M_B$ terms.
Or one may write down all possible next order terms in $1/M_B$
\cite{BKKM92} and then
fix the coefficients by using the so-called ``velocity reparameterization
invariance'' \cite{VRIs}.
The two methods should give the same result.
In this paper, therefore, we use the results of Ref. \cite{GSS88} as
discussed in Ref. \cite{BMi95} for a simple estimate on the $1/M_B$
corrections.%
\footnote{See also Ref. \cite{BKKM92} for a critical review.}
If we consider the one-loop self energy of Fig. \ref{fig:ml}(a)
with the intermediate state baryon mass $M_{B'}$ in a fully
relativistic theory according to Ref. \cite{GSS88}, then we have
\begin{equation}
\delta M_B = \frac{i\beta}{2f^2} \int \frac{d^4 k}{(2\pi)^4}
\frac{\gamma_5 {\!\not\hskip-0.4mm k}
 ( {\not\hskip-0.7mm P} + {\!\not\hskip-0.4mm k} + M_{B'} ) \gamma_5
 {\!\not\hskip-0.4mm k}}
{(k^2 - m_\Pi^2)(2 P \cdot k + k^2 )},
\end{equation}
where $\beta$ stands for an SU(3) Clebsch-Gordan coefficient. By expanding
the loop integral, one would have
\begin{eqnarray}
\delta M_B &=& \frac{\beta}{16\pi f^2} \left[
\frac{M_{B'}^3}{\pi} \left( \frac{1}{\epsilon} - \gamma_E + \ln (4\pi) + 1
- \ln M_{B'}^2 \right) \right.
\nonumber \\ && \mbox{}
+ \frac{M_{B'} m_\Pi^2}{\pi} \left( \frac{1}{\epsilon} - \gamma_E +
 \ln (4\pi) + 2 - \ln M_{B'}^2 \right)
\nonumber \\ && \left. \mbox{}
- m_\Pi^3 \left( 1 - \frac{m_\Pi}{\pi M_{B'}}
 \left[ 1 + \ln \frac{M_{B'}}{m_\pi} \right] + \dots \right)
\right],
\end{eqnarray}
where $\epsilon = d-4$ in dimensional regularization.
The first two terms proportional to $M_{B'}^3$ and $M_{B'} m_\Pi^2$ are
the troublesome terms as noted by Ref. \cite{GSS88}.
The $m_\Pi^3$ term is what we obtained previously, and the $m_\Pi^4/M_{B'}$
term is the $1/M_B$ correction we want.
Here we note that the $1/M_B$ correction terms carry the same Clebsch-Gordan
coefficient as the $m_\Pi^3$ term.
This was verified by explicit computation in Ref. \cite{BM97}.
We use this result for our estimate of the $1/M_B$ corrections,
\begin{equation}
\delta M_B = - \frac{m_\Pi^4}{16 \pi^2 f^2 M_B^0}
\left\{ 1 + \frac12 \ln \frac{m_\Pi^2}{\lambda^2} \right\}
(B| {\cal O}^\Pi |B),
\end{equation}
with ${\cal O}^\Pi$ defined in (\ref{loop0-2}).
We can use $M_B^0 = a_0 N_c$ and note that the order of this $\delta M_B$
is $O(\varepsilon^4)$.

Finally, we get the full one-loop correction to the baryon mass as
\begin{eqnarray}
\delta M_B &=&
\sum_{B'} {\cal J}_2 (\delta M_{B'},m_\Pi) (B| \gamma_{B'}^\Pi |B)
- \frac{m_\Pi^4}{16 \pi^2 f^2 M_B^0}
\left\{ 1 + \frac12 \ln \frac{m_\Pi^2}{\lambda^2} \right\}
(B| {\cal O}^\Pi |B)
\nonumber \\ && \mbox{}
+ \frac{m_\Pi^2}{16\pi^2 f^2} \ln \frac{m_\Pi^2}{\lambda^2}
(B| {\cal P}^\Pi |B)
- \frac{m_\Pi^4}{16\pi^2 f^2} \ln \frac{m_\Pi^2}{\lambda^2}
(B| {\cal Q}^\Pi |B),
\label{finalMB}
\end{eqnarray}
where the operators, ${\cal O}^\Pi$, ${\cal P}^\Pi$ and ${\cal Q}^\Pi$ are
respectively given in Eqs. (\ref{Op-O}), (\ref{Op-P}) and (\ref{Op-Q}), and
${\cal J}_2$ is defined in Eq. (\ref{int-J}).
Note that when we calculate the $\gamma_{B'}^\Pi$ term, we should include the
fictitious intermediate baryon states of spin up to 5/2.
{}From the structure of the operators, we can see that the loop corrections to
the mass relations $(\it M1)$, $(\it M2')$ and $(\it M3')$ come from the
$\gamma_{B'}^\Pi$ and $1/M_B$ terms, and the other terms respect the three
mass relations.
Note also that the leading contribution to $\gamma_{B'}^\Pi$ is
$O(\varepsilon^2)$ while those of ${\cal O}^\Pi$, ${\cal P}^\Pi$ and
${\cal Q}^\Pi$ are $O(\varepsilon^4)$.

\section{Sigma term and strangeness contribution to the nucleon mass}

The pion-nucleon sigma term, defined as
\begin{equation}
\sigma_{\pi N} = \hat{m} \langle p | \bar uu + \bar dd |p \rangle,
\end{equation}
can be computed from the expression of the nucleon mass using the
Feynman-Hellman theorem:
\begin{equation}
\sigma_{\pi N} = \hat{m} \frac{\partial M_N}{\partial \hat{m}}.
\end{equation}
The strange quark contribution to the nucleon mass can be written as
\begin{equation}
\langle p | m_s \bar ss | p \rangle = m_s \frac{\partial M_N}{\partial m_s}.
\end{equation}
Then we can estimate the strange quark matrix element (SME)
$\langle p | m_s \bar ss | p \rangle$ from
the mass formulas derived in the previous Sections.

In this Section, we consider the SME at the tree level.
Up to $O(\varepsilon^1)$, the nucleon mass is written as
\begin{equation}
M_N = a_0 N_c + \frac{3}{N_c} a_1 + N_c m_\pi^2 b_1.
\end{equation}
We find that there is no strange quark contribution to the nucleon
mass at this order:
\begin{eqnarray}
\sigma_{\pi N} &=& N_c m_\pi^2 b_1, \nonumber \\
\langle p | m_s \bar ss | p \rangle &=& 0.
\end{eqnarray}
{}From Table \ref{tab:fit}, we observe
\begin{equation}
a = a_0 N_c + N_c m_\pi^2 b_1 = 1063 \mbox{ MeV}, \qquad
b = \frac{a_1}{N_c} = 26.2 \mbox{ MeV},
\end{equation}
where $a$ and $b$ are defined in Eq. (\ref{massform}).
So using $\sigma_{\pi N} = 45$ MeV \cite{GLS91a}, we can fix the three
parameters as
\begin{equation}
a_0 = 339.3\ \  [337.7] \mbox{ MeV}, \qquad a_1 = 78.6 \mbox{ MeV}, \qquad
b_1 = 7.88 \ \  [8.75] \times 10^{-4} \mbox{ MeV}^{-1},
\end{equation}
where the values in square brackets correspond to
$\sigma_{\pi N} = 50$ MeV as suggested by the lattice calculation of
Ref. \cite{DLL96}.

The non-vanishing SME comes from the $O(\varepsilon^2)$ terms.
The nucleon mass up to this order reads
\begin{equation}
M_N = a_0 N_c + \frac{3}{N_c} a_1 + N_c m_\pi^2 b_1 +
(2 m_K^2 + m_\pi^2) \alpha_1,
\end{equation}
and involves four parameters. We find
\begin{eqnarray}
\sigma_{\pi N} &=& m_\pi^2 (N_c b_1 + 2 \alpha_1), \nonumber \\
\langle p | m_s \bar ss | p \rangle &=& (2 m_K^2 - m_\pi^2) \alpha_1,
\end{eqnarray}
which gives
\begin{equation}
\langle p | m_s \bar ss | p \rangle = \frac{1}{2} ( 2m_K^2 - m_\pi^2)
\left( \frac{\sigma_{\pi N}}{m_\pi^2} - N_c b_1 \right).
\label{sme0}
\end{equation}
Note that the SME starts at $O(N_c^0)$ in $1/N_c$ counting as pointed out
in Ref. \cite{Luty95}.
{}From the best fit of Table \ref{tab:fit}, we get
\begin{eqnarray}
a &=& a_0 N_c + N_c m_\pi^2 b_1 + (2 m_K^2 + m_\pi^2) \alpha_1 =
923.9 \mbox{ MeV}, \nonumber \\
b &=& \frac{a_1}{N_c} = 19.54 \mbox{ MeV}, \nonumber \\
c &=& 2 (m_K^2 - m_\pi^2) b_1 = 159 \mbox{ MeV},
\end{eqnarray}
which gives
\begin{eqnarray}
&& a_0 = 190.45 \ \ [168.05] \mbox{ MeV}, \qquad
a_1 = 58.62 \mbox{ MeV}, \nonumber \\
&& b_1 = 3.52 \times 10^{-4} \mbox{ MeV}^{-1}, \qquad
\alpha_1 = 6.53\ \ [7.85] \times 10^{-4} \mbox{ MeV}^{-1},
\end{eqnarray}
for $\sigma_{\pi N}=45$ MeV (the values in the square brackets are for
$\sigma_{\pi N} = 50$ MeV).
Then we have
\begin{equation}
\langle p | m_s \bar ss | p \rangle = 307.8 \ \ [369.6] \mbox{ MeV}.
\end{equation}
This shows the familiar strong dependence of $\langle p | m_s \bar s s |
p \rangle$ on the value of $\sigma_{\pi N}$.
This is because the constant multiplying $\sigma_{\pi N}$ in Eq. (\ref{sme0})
is as large as $12.4$.
For example, if we use $\sigma_{\pi N} = 65$ MeV,
we find $555$ MeV for the SME.

However, we have to include at least the $O(\varepsilon^3)$ terms to get
a more reliable value of SME because the fitted baryon mass spectra is
reasonably consistent with the experiment from this order onward.
For the nucleon mass we have two additional terms so that
\begin{equation}
M_N = a_0 N_c + \frac{3}{N_c} a_1 + N_c m_\pi^2 b_1 +
(2 m_K^2 + m_\pi^2) \alpha_1 + \frac{3}{N_c} m_\pi^2 b_2 + m_\pi^4 N_c
(c_1+c_2).
\end{equation}
Although there are altogether 7 parameters in the Lagrangian, we have only
6 independent parameters since $c_1$ and $c_2$ enter in the form
$(c_1+c_2)$ for all baryon masses.
The final result is:
\begin{eqnarray}
\sigma_{\pi N} &=& m_\pi^2 [N_c b_1 + 2 \alpha_1 + \frac{3}{N_c} b_2 +
2 N_c m_\pi^2 (c_1+c_2) ], \nonumber \\
\langle p | m_s \bar ss | p \rangle &=& (2 m_K^2 - m_\pi^2) \alpha_1,
\end{eqnarray}
which implies
\begin{equation}
\langle p | m_s \bar ss | p \rangle = \frac{1}{2} ( 2m_K^2 - m_\pi^2)
\left\{ \frac{\sigma_{\pi N}}{m_\pi^2} - N_c b_1 - \frac{3}{N_c} b_2
- 2 N_c m_\pi^2 (c_1+c_2) \right\}.
\label{SME-3}
\end{equation}
To estimate this matrix element, we must determine the parameters.
Not all of them can be fixed, however, since there are 6 parameters
while we have only 5 pieces of information: four from baryon masses and
one from the $\pi N$ sigma term.
{}From Table \ref{tab:fit}, we have
\begin{eqnarray}
a &=& a_0 N_c + N_c m_\pi^2 b_1 + (2m_K^2 + m_\pi^2) \alpha_1 + 
N_c m_\pi^4 (c_1+c_2) = 863.7 \mbox{ MeV}, \nonumber \\
b &=& \frac{a_1}{N_c} + \frac{m_\pi^2}{N_c} b_2 = 25.0 \mbox{ MeV}, \nonumber \\
c &=& 2(m_K^2 - m_\pi^2) b_1 = 227.8 \mbox{ MeV}, \nonumber \\
d &=& \frac{2}{N_c} (m_K^2 - m_\pi^2) b_2 = -16.6 \mbox{ MeV},
\end{eqnarray}
which gives
\begin{equation}
a_1 = 77.03 \mbox{ MeV}, \qquad
b_1 = 5.04 \times 10^{-4} \mbox{ MeV}^{-1}, \qquad
b_2 = -1.10 \times 10^{-4} \mbox{ MeV}^{-1}.
\end{equation}
Note that these best fit values of $a_1$ and $b_1$ at $O(\varepsilon^3)$
are between the values found at $O(\varepsilon^1)$ and at $O(\varepsilon^2)$.
Since the other parameters cannot be determined uniquely, we rewrite
the SME of Eq. (\ref{SME-3}) in the form:
\begin{equation}
\langle p | m_s \bar ss | p \rangle = \frac12 \left( 1 -
\frac{m_\pi^2}{2 m_K^2} \right)
\left\{ 2( a - a_0 N_c) - \sigma_{\pi N} -
m_\pi^2 \left( N_c b_1 - \frac{3}{N_c} b_2 \right) \right\},
\end{equation}
where we have expressed $(c_1+c_2)$ in terms of $\sigma_{\pi N}$ and $a$.
Since $a$ is fixed by the mass spectrum, therefore, the SME of the above form
depends on the {\it unfixed} parameter $a_0$.
For a numerical estimate we can use the fitted values of $a_0$ from the
calculations at $O(\varepsilon^1)$ and $O(\varepsilon^2)$, i.e.,
$a_0 = 190 \sim 340$ MeV.
This leads to $ \langle p | m_s \bar ss | p \rangle$ ranging between
about $250$ MeV and $-190$ MeV.
Now the dependence on the $\pi N$ sigma term is very weak, while it
depends strongly on the value of $a_0$, leaving $\langle p | m_s \bar{s}
s | p \rangle$ almost completely uncertain.

At $O(\varepsilon^4)$ and $O(\varepsilon^5)$, the situation becomes even
more subtle.
There are 9 parameters with 5 pieces of information in case of 
$O(\varepsilon^4)$.
If we take into account the corrections from $O(\varepsilon^5)$, then we
have 13 effective parameters%
\footnote{There are totally 15 parameters up to $O(\varepsilon^5)$ calculation.
However, the three parameters $d_{1,2,3}$ appear only in the form of
$(d_1+d_2+d_3)$ at this order. So there are effectively 13 parameters.}
with 7 constraints.
Additional information is therefore required such as isospin symmetry
breaking effects in the baryon spectra and/or $KN$ sigma terms \cite{BKM93}.
As a reference, we give the formulas of the sigma term and the SME up to
$O(\varepsilon^4)$ below:
\begin{eqnarray}
\sigma_{\pi N} &=& m_\pi^2 \left\{ N_c b_1 + 2 \alpha_1 + \frac{3}{N_c} b_2
+ 2 N_c m_\pi^2 (c_1+c_2) + \frac{6}{N_c^2} \alpha_2 + 4 m_\pi^2 \beta_1
\right\}, \nonumber \\
\langle p | m_s \bar ss | p \rangle &=& ( 2m_K^2 - m_\pi^2)
\left\{ \alpha_1 + \frac{3}{N_c^2} \alpha_2 + 2(2m_K^2-m_\pi^2) \beta_1
\right\},
\end{eqnarray}
where
\begin{eqnarray}
a &=& a_0 N_c + N_c m_\pi^2 b_1 + (2m_K^2 + m_\pi^2) \alpha_1 + 
      N_c m_\pi^4 (c_1+c_2)
\nonumber \\ && \mbox{}
+ (4m_K^2 - 4m_K^2 m_\pi^2 + 3 m_\pi^4) \beta_1
= 863.7 \mbox{ MeV}, \nonumber \\
b &=& \frac{a_1}{N_c} + \frac{m_\pi^2}{N_c} b_2
+ \frac{1}{N_c^2} ( 2m_K^2 + m_\pi^2) \alpha_2 = 25.0 \mbox{ MeV}, \nonumber \\
c &=& 2(m_K^2 - m_\pi^2) b_1 + 4(m_K^2-m_\pi^2) [ m_K^2 c_1 + m_\pi^2 c_2 ]
= 227.8 \mbox{ MeV}, \nonumber \\
d &=& \frac{2}{N_c} (m_K^2 - m_\pi^2) b_2 = -16.6 \mbox{ MeV}.
\end{eqnarray}
In essence one observes that corrections beyond the standard estimate
(\ref{sme0}) for $\langle p| m_s \bar{s} s |p \rangle$ are so large that
they prohibit quantitative conclusions about the strange quark
contribution to the nucleon mass.

\section{Summary and Discussion}

In summary, we have re-analyzed baryon masses within baryon chiral
perturbation theory in combination with the large $N_c$ expansion.
Before computing the baryon masses, we have calculated the baryon
axial current.
We find that the two diagrams of Fig. \ref{fig:ax1} give contributions
of the same order in $1/N_c$ counting.
Inclusion of the wave function renormalization terms is crucial to get the
right order for the one-loop corrections because this gives the proper
commutator structure to the baryon axial current operator.
However, when calculating  $\alpha^8_{B'B}$, two-body
operators give contributions of the same order as one-body operators.
Unless the coupling constants of the general $n$-body operators
are suppressed numerically, their effects must be included in order to be
consistent with the $1/N_c$ expansion.

Next, we have considered the baryon mass spectrum in this scheme.
For this aim, we have used that both $m_\Pi$ and $1/\delta M$ scale as
$O(\varepsilon)$, where $m_\Pi$ and $\delta M$, respectively, represent the
Goldstone boson mass and the octet-decuplet mass splitting which depends
on $1/N_c$.
At the tree level, we found that the empirical mass spectrum is well
reproduced to $O(\varepsilon^3)$ and the corrections from
$O(\varepsilon^5)$ are not so crucial. But the Gell-Mann - Okubo
mass relation and the
equal spacing rule in the decuplet are modified at $O(\varepsilon^5)$.
At the one-loop level, there is no additional contribution that
gives the characteristic commutator structure, and the loop corrections
seem to violate the $1/N_c$ expansion.
However, the leading terms are constant for all baryon states and can be
safely absorbed into the central baryon mass in the chiral limit.
The meson loop corrections involving the operators ${\cal O}^\Pi$ with
the coupling constant $g$, ${\cal P}^\Pi$ and ${\cal Q}^\Pi$ of
Eq. (\ref{finalMB}) satisfy the modified mass relations $(\it M1)$,
$(\it M2')$ and $(\it M3')$.
To get the correct result, the intermediate baryon states must
include fictitious states of spin up to 5/2 in order to satisfy
the closure relation, $\sum_{B} |B) (B| = 1$, for the spin-1 operator
$X^{ia}$ in the large $N_c$ limit.
As in the calculation of $\alpha^8_{B'B}$, the $\eta$-meson loop
corrections to the baryon self energy require general $n$-body operators in
order to be consistent with the $1/N_c$ expansion.

Finally we have estimated the strangeness contribution to the nucleon mass
at the tree level.
We confirmed that this matrix element is $O(N_c^0)$ in the
$1/N_c$ counting.
At leading order, namely $O(\varepsilon^2)$, this contribution can amount
to more than 300 MeV.
At the next order, we cannot uniquely determine the mass parameters
because of lack of independent empirical information.
But the upper bound of the strangeness contribution to the nucleon mass
is now reduced to around 250 MeV.

\acknowledgements

We thank N. Kaiser for useful discussions.
One of us (Y.O.) acknowledges the financial support from the Alexander
von Humboldt Foundation.
This work was supported in part by the Korea Science and Engineering
Foundation through CTP of Seoul National University.


\appendix

\section{Baryon States}

The octet and decuplet baryons states $|B, j_z)$ in the number space are
given in this Appendix. For simplicity we give only the $s_z=+1/2$ states
for baryon octet and $s_z = +3/2$ states for baryon decuplet.
Other spin states can be obtained straightforwardly.
The octet states are
\begin{eqnarray}
|p, {\textstyle +\frac12} ) &=&
C_N \, \alpha_1^\dagger\,
(A_s^\dagger)^n |0),
\nonumber \\
|n, {\textstyle +\frac12} ) &=&
C_N \, \alpha_3^\dagger\,
(A_s^\dagger)^n |0),
\\
|\Lambda, {\textstyle +\frac12} ) &=&
- C_\Lambda \, \alpha_5^\dagger\,
(A_s^\dagger)^n |0),
\\
|\Sigma^+, {\textstyle +\frac12} ) &=&
- C_\Sigma \, \alpha_1^\dagger\,
(A_d^\dagger) \,
(A_s^\dagger)^{n-1} |0),
\nonumber \\
|\Sigma^0, {\textstyle +\frac12} ) &=&
\frac{1}{\sqrt{2}}\, C_\Sigma \, \{
\alpha_1^\dagger\, A_u^\dagger + \alpha_3^\dagger\, A_d^\dagger \} \,
(A_s^\dagger)^{n-1} |0),
\nonumber \\
|\Sigma^-, {\textstyle +\frac12} ) &=&
C_\Sigma \, \alpha_3^\dagger\, A_u^\dagger \,
(A_s^\dagger)^{n-1} |0),
\\
|\Xi^0, {\textstyle +\frac12} ) &=&
- C_\Xi \, \alpha_5^\dagger\,
A_d^\dagger \,
(A_s^\dagger)^{n-1} |0),
\nonumber \\
|\Xi^-, {\textstyle +\frac12} ) &=&
C_\Xi \, \alpha_5^\dagger\, A_u^\dagger \,
(A_s^\dagger)^{n-1} |0),
\end{eqnarray}
where $A_u^\dagger = \alpha_3^\dagger \alpha_6^\dagger - \alpha_4^\dagger
\alpha_5^\dagger$,
$A_d^\dagger = \alpha_1^\dagger \alpha_6^\dagger - \alpha_2^\dagger
\alpha_5^\dagger$ and
$A_s^\dagger = \alpha_1^\dagger \alpha_4^\dagger - \alpha_2^\dagger
\alpha_3^\dagger$,
with the normalization constants
\begin{eqnarray}
\begin{array}{ll}
\displaystyle
\left[ n! C_N \right]^2 = \frac{2}{(n+1)(n+2)}, \qquad &
\displaystyle
\left[ n! C_\Lambda \right]^2 = \frac{1}{n+1}, \\[4mm]
\displaystyle
\left[ (n-1)! C_\Sigma \right]^2 = \frac{2}{n(n+1)(n+2)}, \qquad &
\displaystyle
\left[ (n-1)! C_\Xi \right]^2 = \frac{2}{3n(n+1)}, 
\end{array}
\end{eqnarray}
from the condition $(B, j_z | B, j_z) = 1$, where $N_c = 2n+1$.
The negative signs of some states were introduced to be
consistent with the quark model convention \cite{Close}.
Explicitly the spin-up proton state can be written as
\begin{equation}
|p, + {\textstyle \frac12}) = C_N \sum_{k=0}^n \frac{n!}{k! (n-k)!} (-1)^k
| n-k+1, k, k, n-k, 0, 0),
\end{equation}
by making use of
\begin{equation}
(A+B)^n = \sum_{k=0}^n \frac{n!}{k!(n-k)!} A^{n-k} B^k.
\end{equation}

The decuplet states are as follows:
\begin{eqnarray}
|\Delta^{++}, {\textstyle +\frac32} ) &=&
C_\Delta \, \alpha_1^\dagger \, \alpha_1^\dagger \, \alpha_1^\dagger \,
(A_s^\dagger)^{n-1} |0),
\nonumber \\
|\Delta^{+}, {\textstyle +\frac32} ) &=&
\sqrt{3}\, C_\Delta \,
\alpha_1^\dagger\, \alpha_1^\dagger \, \alpha_3^\dagger \,
(A_s^\dagger)^{n-1} |0),
\nonumber \\
|\Delta^{0}, {\textstyle +\frac32} ) &=&
\sqrt{3}\, C_\Delta \,
\alpha_1^\dagger\, \alpha_3^\dagger \, \alpha_3^\dagger \,
(A_s^\dagger)^{n-1} |0),
\nonumber \\
|\Delta^{-}, {\textstyle +\frac32} ) &=&
C_\Delta \, \alpha_3^\dagger \, \alpha_3^\dagger \, \alpha_3^\dagger \,
(A_s^\dagger)^{n-1} |0),
\\
|\Sigma^{*+}, {\textstyle +\frac32} ) &=&
C_{\Sigma^*} \, \alpha_1^\dagger\, \alpha_1^\dagger \, \alpha_5^\dagger \,
(A_s^\dagger)^{n-1} |0),
\nonumber \\
|\Sigma^{*0}, {\textstyle +\frac32} ) &=&
\sqrt{2}\, C_{\Sigma^*} \,
\alpha_1^\dagger \, \alpha_3^\dagger \, \alpha_5^\dagger \,
(A_s^\dagger)^{n-1} |0),
\nonumber \\
|\Sigma^{*-}, {\textstyle +\frac32} ) &=&
C_{\Sigma^*} \, \alpha_3^\dagger \, \alpha_3^\dagger \, \alpha_5^\dagger \,
(A_s^\dagger)^{n-1} |0),
\\
|\Xi^{*0}, {\textstyle +\frac32} ) &=&
C_{\Xi^*} \, \alpha_1^\dagger\, \alpha_5^\dagger \, \alpha_5^\dagger \,
(A_s^\dagger)^{n-1} |0),
\nonumber \\
|\Xi^{*-}, {\textstyle +\frac32} ) &=&
C_{\Xi^*} \, \alpha_3^\dagger\, \alpha_5^\dagger \, \alpha_5^\dagger \,
(A_s^\dagger)^{n-1} |0),
\\
|\Omega, {\textstyle +\frac32} ) &=&
C_\Omega^* \, \alpha_5^\dagger\, \alpha_5^\dagger \, \alpha_5^\dagger \,
(A_s^\dagger)^{n-1} |0),
\end{eqnarray}
where
\begin{eqnarray}
\begin{array}{ll}
\displaystyle
\left[ (n-1)! C_\Delta \right]^2 = \frac{4}{n(n+1)(n+2)(n+3)}, \qquad &
\displaystyle
\left[ (n-1)! C_{\Sigma^*} \right]^2 = \frac{3}{n(n+1)(n+2)}, \\[4mm]
\displaystyle
\left[ (n-1)! C_{\Xi^*} \right]^2 = \frac{1}{n(n+1)}, \qquad &
\displaystyle
\left[ (n-1)! C_\Omega \right]^2 = \frac{1}{6n}.
\end{array}
\end{eqnarray}

However, the octet and decuplet states are not sufficient to satisfy the
closure relation (\ref{closure}) in the large $N_c$ limit, and we have
to include higher spin states.
Since $X^{ia}$ is a spin-1 operator, it is sufficient to introduce
fictitious states up to spin 5/2.
These states are distinguished from the octet/decuplet by a tilde and we
denote the strangeness ${\cal S}=-4$ states by $|{\rm S})$.
These states can be obtained by considering 5-quark states
multiplied by $(A_s^\dagger)^{n-2}$, whereas the conventional octet and
decuplet are formed by 3-quark states with $(A_s^\dagger)^{n-1}$.
Then the fictitious states of spin 1/2 are
\begin{eqnarray}
| \tilde \Xi_1, {\textstyle +\frac12} ) &=&
C_{\tilde \Xi}\, \alpha_1^\dagger \, A_d^\dagger \, A_d^\dagger \,
(A_s^\dagger)^{n-2} |0),
\nonumber \\
| \tilde \Xi_2, {\textstyle +\frac12} ) &=&
\frac{1}{\sqrt{3}}\, C_{\tilde \Xi}\, \left\{
2 \alpha_1^\dagger \, A_u^\dagger
+ \alpha_3^\dagger \, A_d^\dagger
\right\}
A_d^\dagger \,
(A_s^\dagger)^{n-2} |0),
\nonumber \\
| \tilde \Xi_3, {\textstyle +\frac12} ) &=&
\frac{1}{\sqrt{3}}\, C_{\tilde \Xi}\, \left\{
\alpha_1^\dagger \, A_u^\dagger + 2 \alpha_3^\dagger \, A_d^\dagger \right\}
A_u^\dagger \,
(A_s^\dagger)^{n-2} |0),
\nonumber \\
| \tilde \Xi_4, {\textstyle +\frac12} ) &=&
C_{\tilde \Xi}\, \alpha_3^\dagger \, A_u^\dagger \, A_u^\dagger \,
(A_s^\dagger)^{n-2} |0),
\\
| \tilde \Omega_1, {\textstyle +\frac12} ) &=&
C_{\tilde \Omega}\, \alpha_5^\dagger \, A_d^\dagger \, A_d^\dagger \,
(A_s^\dagger)^{n-2} |0),
\nonumber \\
| \tilde \Omega_2, {\textstyle +\frac12} ) &=&
\sqrt{2}\, C_{\tilde \Omega}\, \alpha_5^\dagger \, A_u^\dagger \, A_d^\dagger \,
(A_s^\dagger)^{n-2} |0),
\nonumber \\
| \tilde \Omega_3, {\textstyle +\frac12} ) &=&
C_{\tilde \Omega}\, \alpha_5^\dagger \, A_u^\dagger \, A_u^\dagger \,
(A_s^\dagger)^{n-2} |0),
\end{eqnarray}
where
\begin{eqnarray}
\left[ (n-2)! C_{\tilde \Xi} \right]^2 &=& \frac{1}{(n-1)n(n+1)(n+2)},
\nonumber \\
\left[ (n-2)! C_{\tilde \Omega} \right]^2 &=& \frac{1}{4(n-1)n(n+1)}.
\end{eqnarray}

For the spin 3/2 states, we have
\begin{eqnarray}
| \tilde \Sigma^*_1, {\textstyle +\frac32} ) &=&
C_{\tilde \Sigma^*}\,
\alpha_1^\dagger \, \alpha_1^\dagger \, \alpha_1^\dagger \, A_d^\dagger
(A_s^\dagger)^{n-2} |0),
\nonumber \\
| \tilde \Sigma^*_2, {\textstyle +\frac32} ) &=&
\frac{1}{2}\, C_{\tilde \Sigma^*}\, \left\{
\alpha_1^\dagger \, \alpha_1^\dagger \, \alpha_1^\dagger \, A_u^\dagger
+ 3 \alpha_1^\dagger \, \alpha_1^\dagger \, \alpha_3^\dagger \, A_d^\dagger
\right\}
(A_s^\dagger)^{n-2} |0),
\nonumber \\
| \tilde \Sigma^*_3, {\textstyle +\frac32} ) &=&
\sqrt{\frac{3}{2}}\, C_{\tilde \Sigma^*}\, \left\{
\alpha_1^\dagger \, \alpha_1^\dagger \, \alpha_3^\dagger \, A_u^\dagger
+ \alpha_1^\dagger \, \alpha_3^\dagger \, \alpha_3^\dagger \, A_d^\dagger
\right\}
(A_s^\dagger)^{n-2} |0),
\nonumber \\
| \tilde \Sigma^*_4, {\textstyle +\frac32} ) &=&
\frac{1}{2}\, C_{\tilde \Sigma^*}\, \left\{
3 \alpha_1^\dagger \, \alpha_3^\dagger \, \alpha_3^\dagger \, A_u^\dagger
+ \alpha_3^\dagger \, \alpha_3^\dagger \, \alpha_3^\dagger \, A_d^\dagger
\right\}
(A_s^\dagger)^{n-2} |0),
\nonumber \\
| \tilde \Sigma^*_5, {\textstyle +\frac32} ) &=&
C_{\tilde \Sigma^*}\,
\alpha_3^\dagger \, \alpha_3^\dagger \, \alpha_3^\dagger \, A_u^\dagger
(A_s^\dagger)^{n-2} |0),
\\
| \tilde \Xi^*_1, {\textstyle +\frac32} ) &=&
C_{\tilde \Xi^*}\,
\alpha_1^\dagger \, \alpha_1^\dagger \, \alpha_5^\dagger \, A_d^\dagger
(A_s^\dagger)^{n-2} |0),
\nonumber \\
| \tilde \Xi^*_2, {\textstyle +\frac32} ) &=&
\frac{1}{\sqrt{3}}\, C_{\tilde \Xi^*}\, \left\{
\alpha_1^\dagger \, \alpha_1^\dagger \, \alpha_5^\dagger \, A_u^\dagger
+ 2\alpha_1^\dagger \, \alpha_3^\dagger \, \alpha_5^\dagger \, A_d^\dagger
\right\}
(A_s^\dagger)^{n-2} |0),
\nonumber \\
| \tilde \Xi^*_3, {\textstyle +\frac32} ) &=&
\frac{1}{\sqrt{3}}\, C_{\tilde \Xi^*}\, \left\{
2 \alpha_1^\dagger \, \alpha_3^\dagger \, \alpha_5^\dagger \, A_u^\dagger
+ \alpha_3^\dagger \, \alpha_3^\dagger \, \alpha_5^\dagger \, A_d^\dagger
\right\}
(A_s^\dagger)^{n-2} |0),
\nonumber \\
| \tilde \Xi^*_4, {\textstyle +\frac32} ) &=&
C_{\tilde \Xi^*}\,
\alpha_3^\dagger \, \alpha_3^\dagger \, \alpha_5^\dagger \, A_u^\dagger
(A_s^\dagger)^{n-2} |0),
\\
| \tilde \Omega^*_1, {\textstyle +\frac32} ) &=&
C_{\tilde \Omega^*}\,
\alpha_1^\dagger \, \alpha_5^\dagger \, \alpha_5^\dagger \, A_d^\dagger
(A_s^\dagger)^{n-2} |0),
\nonumber \\
| \tilde \Omega^*_2, {\textstyle +\frac32} ) &=&
\frac{1}{\sqrt{2}}\, C_{\tilde \Omega^*}\, \left\{
\alpha_1^\dagger \, \alpha_5^\dagger \, \alpha_5^\dagger \, A_u^\dagger
+ \alpha_3^\dagger \, \alpha_5^\dagger \, \alpha_5^\dagger \, A_d^\dagger
\right\}
(A_s^\dagger)^{n-2} |0),
\nonumber \\
| \tilde \Omega^*_3, {\textstyle +\frac32} ) &=&
C_{\tilde \Omega^*}\,
\alpha_3^\dagger \, \alpha_5^\dagger \, \alpha_5^\dagger \, A_u^\dagger
(A_s^\dagger)^{n-2} |0),
\\
| \tilde {\rm S}^*_1, {\textstyle +\frac32} ) &=&
C_{\tilde {\rm S}^*}\,
\alpha_5^\dagger \, \alpha_5^\dagger \, \alpha_5^\dagger \, A_d^\dagger
(A_s^\dagger)^{n-2} |0),
\nonumber \\
| \tilde {\rm S}^*_2, {\textstyle +\frac32} ) &=&
C_{\tilde {\rm S}^*}\,
\alpha_5^\dagger \, \alpha_5^\dagger \, \alpha_5^\dagger \, A_u^\dagger
(A_s^\dagger)^{n-2} |0),
\end{eqnarray}
where
\begin{eqnarray}
\left[ (n-2)! C_{\tilde \Sigma^*} \right]^2 &=&
\frac{4}{(n-1)n(n+1)(n+2)(n+3)},
\nonumber \\
\left[ (n-2)! C_{\tilde \Xi^*} \right]^2 &=&
\frac{12}{5(n-1)n(n+1)(n+2)},
\nonumber \\
\left[ (n-2)! C_{\tilde \Omega^*} \right]^2 &=&
\frac{3}{5 (n-1)n(n+1)},
\nonumber \\
\left[ (n-2)! C_{\tilde {\rm S}^*} \right]^2 &=&
\frac{1}{15 (n-1)n}.
\end{eqnarray}

The possible spin 5/2 states are
\begin{eqnarray}
| \tilde \Delta^{**}_1, {\textstyle +\frac52} ) &=&
C_{\tilde \Delta^{**}}\,
\alpha_1^\dagger \, \alpha_1^\dagger \, \alpha_1^\dagger \,
\alpha_1^\dagger \, \alpha_1^\dagger \,
(A_s^\dagger)^{n-2} |0),
\nonumber \\
| \tilde \Delta^{**}_2, {\textstyle +\frac52} ) &=&
\sqrt{5}\, C_{\tilde \Delta^{**}}\,
\alpha_1^\dagger \, \alpha_1^\dagger \, \alpha_1^\dagger \,
\alpha_1^\dagger \, \alpha_3^\dagger \,
(A_s^\dagger)^{n-2} |0),
\nonumber \\
| \tilde \Delta^{**}_3, {\textstyle +\frac52} ) &=&
\sqrt{5}\, C_{\tilde \Delta^{**}}\,
\alpha_1^\dagger \, \alpha_1^\dagger \, \alpha_1^\dagger \,
\alpha_3^\dagger \, \alpha_3^\dagger \,
(A_s^\dagger)^{n-2} |0),
\nonumber \\
| \tilde \Delta^{**}_4, {\textstyle +\frac52} ) &=&
\sqrt{10}\, C_{\tilde \Delta^{**}}\,
\alpha_1^\dagger \, \alpha_1^\dagger \, \alpha_3^\dagger \,
\alpha_3^\dagger \, \alpha_3^\dagger \,
(A_s^\dagger)^{n-2} |0),
\nonumber \\
| \tilde \Delta^{**}_5, {\textstyle +\frac52} ) &=&
\sqrt{5}\, C_{\tilde \Delta^{**}}\,
\alpha_1^\dagger \, \alpha_3^\dagger \, \alpha_3^\dagger \,
\alpha_3^\dagger \, \alpha_3^\dagger \,
(A_s^\dagger)^{n-2} |0),
\nonumber \\
| \tilde \Delta^{**}_6, {\textstyle +\frac52} ) &=&
C_{\tilde \Delta^{**}}\,
\alpha_3^\dagger \, \alpha_3^\dagger \, \alpha_3^\dagger \,
\alpha_3^\dagger \, \alpha_3^\dagger \,
(A_s^\dagger)^{n-2} |0),
\\
| \tilde \Sigma^{**}_1, {\textstyle +\frac52} ) &=&
C_{\tilde \Sigma^{**}}\,
\alpha_1^\dagger \, \alpha_1^\dagger \, \alpha_1^\dagger \,
\alpha_1^\dagger \, \alpha_5^\dagger \,
(A_s^\dagger)^{n-2} |0),
\nonumber \\
| \tilde \Sigma^{**}_2, {\textstyle +\frac52} ) &=&
2\, C_{\tilde \Sigma^{**}}\,
\alpha_1^\dagger \, \alpha_1^\dagger \, \alpha_1^\dagger \,
\alpha_3^\dagger \, \alpha_5^\dagger \,
(A_s^\dagger)^{n-2} |0),
\nonumber \\
| \tilde \Sigma^{**}_3, {\textstyle +\frac52} ) &=&
\sqrt{6} \, C_{\tilde \Sigma^{**}}\,
\alpha_1^\dagger \, \alpha_1^\dagger \, \alpha_3^\dagger \,
\alpha_3^\dagger \, \alpha_5^\dagger \,
(A_s^\dagger)^{n-2} |0),
\nonumber \\
| \tilde \Sigma^{**}_4, {\textstyle +\frac52} ) &=&
2 \, C_{\tilde \Sigma^{**}}\,
\alpha_1^\dagger \, \alpha_3^\dagger \, \alpha_3^\dagger \,
\alpha_3^\dagger \, \alpha_5^\dagger \,
(A_s^\dagger)^{n-2} |0),
\nonumber \\
| \tilde \Sigma^{**}_5, {\textstyle +\frac52} ) &=&
C_{\tilde \Sigma^{**}}\,
\alpha_3^\dagger \, \alpha_3^\dagger \, \alpha_3^\dagger \,
\alpha_3^\dagger \, \alpha_5^\dagger \,
(A_s^\dagger)^{n-2} |0),
\\
| \tilde \Xi^{**}_1, {\textstyle +\frac52} ) &=&
C_{\tilde \Xi^{**}}\,
\alpha_1^\dagger \, \alpha_1^\dagger \, \alpha_1^\dagger \,
\alpha_5^\dagger \, \alpha_5^\dagger \,
(A_s^\dagger)^{n-2} |0),
\nonumber \\
| \tilde \Xi^{**}_2, {\textstyle +\frac52} ) &=&
\sqrt{3} \, C_{\tilde \Xi^{**}}\,
\alpha_1^\dagger \, \alpha_1^\dagger \, \alpha_3^\dagger \,
\alpha_5^\dagger \, \alpha_5^\dagger \,
(A_s^\dagger)^{n-2} |0),
\nonumber \\
| \tilde \Xi^{**}_3, {\textstyle +\frac52} ) &=&
\sqrt{3} \, C_{\tilde \Xi^{**}}\,
\alpha_1^\dagger \, \alpha_3^\dagger \, \alpha_3^\dagger \,
\alpha_5^\dagger \, \alpha_5^\dagger \,
(A_s^\dagger)^{n-2} |0),
\nonumber \\
| \tilde \Xi^{**}_4, {\textstyle +\frac52} ) &=&
C_{\tilde \Xi^{**}}\,
\alpha_3^\dagger \, \alpha_3^\dagger \, \alpha_3^\dagger \,
\alpha_5^\dagger \, \alpha_5^\dagger \,
(A_s^\dagger)^{n-2} |0),
\\
| \tilde \Omega^{**}_1, {\textstyle +\frac52} ) &=&
C_{\tilde \Omega^{**}}\,
\alpha_1^\dagger \, \alpha_1^\dagger \, \alpha_5^\dagger \,
\alpha_5^\dagger \, \alpha_5^\dagger \,
(A_s^\dagger)^{n-2} |0),
\nonumber \\
| \tilde \Omega^{**}_2, {\textstyle +\frac52} ) &=&
\sqrt{2} \, C_{\tilde \Omega^{**}}\,
\alpha_1^\dagger \, \alpha_3^\dagger \, \alpha_5^\dagger \,
\alpha_5^\dagger \, \alpha_5^\dagger \,
(A_s^\dagger)^{n-2} |0),
\nonumber \\
| \tilde \Omega^{**}_3, {\textstyle +\frac52} ) &=&
C_{\tilde \Omega^{**}}\,
\alpha_3^\dagger \, \alpha_3^\dagger \, \alpha_5^\dagger \,
\alpha_5^\dagger \, \alpha_5^\dagger \,
(A_s^\dagger)^{n-2} |0),
\\
| \tilde {\rm S}^{**}_1, {\textstyle +\frac52} ) &=&
C_{\tilde {\rm S}^{**}}\,
\alpha_1^\dagger \, \alpha_5^\dagger \, \alpha_5^\dagger \,
\alpha_5^\dagger \, \alpha_5^\dagger \,
(A_s^\dagger)^{n-2} |0),
\nonumber \\
| \tilde {\rm S}^{**}_2, {\textstyle +\frac52} ) &=&
C_{\tilde {\rm S}^{**}}\,
\alpha_3^\dagger \, \alpha_5^\dagger \, \alpha_5^\dagger \,
\alpha_5^\dagger \, \alpha_5^\dagger \,
(A_s^\dagger)^{n-2} |0),
\end{eqnarray}
where
\begin{eqnarray}
\left[ (n-2)! C_{\tilde \Delta^{**}} \right]^2 &=&
\frac{6}{(n-1)n(n+1)(n+2)(n+3)(n+4)},
\nonumber \\
\left[ (n-2)! C_{\tilde \Sigma^{**}} \right]^2 &=&
\frac{5}{(n-1)n(n+1)(n+2)(n+3)},
\nonumber \\
\left[ (n-2)! C_{\tilde \Xi^{**}} \right]^2 &=&
\frac{2}{(n-1)n(n+1)(n+2)},
\nonumber \\
\left[ (n-2)! C_{\tilde \Omega^{**}} \right]^2 &=&
\frac{1}{2(n-1)n(n+1)},
\nonumber \\
\left[ (n-2)! C_{\tilde {\rm S}^{**}} \right]^2 &=&
\frac{1}{12(n-1)n}.
\end{eqnarray}

Note that the $\tilde \Delta$, $\tilde \Sigma$, $\tilde \Xi$, $\tilde\Omega$
and $\tilde {\rm S}$ families have isospin 5/2, 2, 3/2, 1 and 1/2, respectively,
and their normalization constants contain the factor $1/(n-1)$ so that
these states can not be defined with $N_c=3$.
Using the results given in Appendix C, one can find that these fictitious
states ensure the relation (\ref{closure}).

\section{Explicit results of $\beta_{BB'}^{\lowercase{i},\Pi}$}

In this Appendix, we give the explicit results of $\beta_{BB'}^{i,\Pi}$
and $\tilde\beta_{BB'}^{i,\Pi}$ of (\ref{beta}) from the $g$ term of
Eq. (\ref{Lag:mb}):

\begin{eqnarray}
\beta_{pn}^{1+i2,\pi} &=&
- \frac23 (N_c+2) g^3 - \frac13 (N_c+2) g, \nonumber \\
\beta_{pn}^{1+i2,K} &=&
- \frac12 (N_c+2) g^3 - \frac16 (N_c+2) g, \nonumber \\
\beta_{pn}^{1+i2,\eta} &=& - \frac{1}{9} (N_c+2) g^3,
\\
\beta_{\Lambda \Sigma^-}^{1+i2,\pi} &=&
- \frac{2}{3\sqrt2} \sqrt{(N_c-1)(N_c+3)} g^3
- \frac{1}{3\sqrt2} \sqrt{(N_c-1)(N_c+3)} g, \nonumber \\
\beta_{\Lambda \Sigma^-}^{1+i2,K} &=&
- \frac{1}{2\sqrt2} \sqrt{(N_c-1)(N_c+3)} g^3
- \frac{1}{6\sqrt2} \sqrt{(N_c-1)(N_c+3)} g, \nonumber \\
\beta_{\Lambda \Sigma^-}^{1+i2,\eta} &=&
- \frac{1}{9\sqrt2} \sqrt{(N_c-1)(N_c+3)} g^3,
\\
\beta_{\Xi^0 \Xi^-}^{1+i2,\pi} &=&
- \frac{2N_c}{9} g^3 - \frac{N_c}{9} g, \nonumber \\
\beta_{\Xi^0 \Xi^-}^{1+i2,K} &=&
- \frac{N_c}{6} g^3 - \frac{N_c}{18} g, \nonumber \\
\beta_{\Xi^0 \Xi^-}^{1+i2,\eta} &=&
- \frac{N_c}{27} g^3,
\\
\beta_{\Sigma^0 \Sigma^-}^{1+i2,\pi} &=&
- \frac{2}{3\sqrt2} (N_c+1) g^3 - \frac{1}{3\sqrt2} (N_c+1) g,
\nonumber \\
\beta_{\Sigma^0 \Sigma^-}^{1+i2,K} &=&
- \frac{1}{2\sqrt2} (N_c+1) g^3 - \frac{1}{6\sqrt2} (N_c+1) g,
\nonumber \\
\beta_{\Sigma^0 \Sigma^-}^{1+i2,\eta} &=&
- \frac{1}{9\sqrt2} (N_c+1) g^3,
\end{eqnarray}
and
\begin{eqnarray}
\beta_{p \Lambda}^{4+i5,\pi} &=& \frac{9}{16} \sqrt{N_c+3} g^3
+ \frac{3}{16} \sqrt{N_c+3} g, \nonumber \\
\beta_{p \Lambda}^{4+i5,K} &=& 2 \beta_{p \Lambda}^{4+i5,\pi}, \nonumber \\
\beta_{p \Lambda}^{4+i5,\eta} &=& \frac{11}{48} \sqrt{N_c+3} g^3
+ \frac{3}{16} \sqrt{N_c+3} g,
\\
\beta_{\Lambda \Xi^-}^{4+i5,\pi} &=& - \frac{9}{16\sqrt3} \sqrt{N_c-1} g^3
- \frac{1}{16} \sqrt{3(N_c-1)} g, \nonumber \\
\beta_{\Lambda \Xi^-}^{4+i5,K} &=& 2 \beta_{p \Lambda}^{4+i5,\pi}, \nonumber \\
\beta_{\Lambda \Xi^-}^{4+i5,\eta} &=& - \frac{11}{48\sqrt3} \sqrt{N_c-1} g^3
- \frac{1}{16} \sqrt{3(N_c-1)} g,
\\
\beta_{p \Sigma^0}^{4+i5,\pi} &=& - \frac{3}{16} \sqrt{N_c-1} g^3
- \frac{1}{16} \sqrt{N_c-1} g, \nonumber \\
\beta_{p \Sigma^0}^{4+i5,K} &=& 2 \beta_{p \Lambda}^{4+i5,\pi}, \nonumber \\
\beta_{p \Sigma^0}^{4+i5,\eta} &=& - \frac{11}{144} \sqrt{N_c-1} g^3
- \frac{1}{16} \sqrt{N_c-1} g,
\\
\beta_{\Sigma^0 \Xi^-}^{4+i5,\pi} &=& - \frac{5\sqrt3}{16} \sqrt{N_c+3} g^3
- \frac{5\sqrt3}{48} \sqrt{N_c+3} g, \nonumber \\
\beta_{\Sigma^0 \Xi^-}^{4+i5,K} &=& 2 \beta_{p \Lambda}^{4+i5,\pi}, \nonumber \\
\beta_{\Sigma^0 \Xi^-}^{4+i5,\eta} &=& - \frac{55\sqrt3}{432} \sqrt{N_c+3} g^3
- \frac{5\sqrt3}{48} \sqrt{N_c+3} g.
\end{eqnarray}
For $\beta^8$, we have
\begin{eqnarray}
\renewcommand{\arraystretch}{2.0}
\begin{array}{lll}
\displaystyle
\beta_{pp}^{8,\pi} = - \frac{3}{2\sqrt3} g^3, \qquad &
\displaystyle
\beta_{pp}^{8,K} = - \frac{1}{4\sqrt3} g^3 - \frac{3}{4\sqrt3} g, \qquad &
\displaystyle
\beta_{pp}^{8,\eta} = - \frac{1}{6\sqrt3} g^3, \\
\displaystyle
\beta_{\Lambda\Lambda}^{8,\pi} = 0, &
\displaystyle
\beta_{\Lambda\Lambda}^{8,K} = \frac{5}{2\sqrt3} g^3
+ \frac{3}{2\sqrt3} g, &
\displaystyle
\beta_{\Lambda\Lambda}^{8,\eta} = \frac{4}{3\sqrt3} g^3, \\
\displaystyle
\beta_{\Sigma\Sigma}^{8,\pi} = - \frac{2}{\sqrt3} g^3, &
\displaystyle
\beta_{\Sigma\Sigma}^{8,K} = - \frac{7}{6\sqrt3} g^3
- \frac{3}{2\sqrt3} g, &
\displaystyle
\beta_{\Sigma\Sigma}^{8,\eta} = -\frac{2}{3\sqrt3} g^3, \\
\displaystyle
\beta_{\Xi\Xi}^{8,\pi} = \frac{2}{\sqrt3} g^3, &
\displaystyle
\beta_{\Xi\Xi}^{8,K} = \frac{41}{12\sqrt3} g^3
+ \frac{9}{4\sqrt3} g, &
\displaystyle
\beta_{\Xi\Xi}^{8,\eta} = \frac{11}{6\sqrt3} g^3.
\end{array}
\end{eqnarray}

The constants $\tilde\beta_{B'B}^{i,\Pi}$ are the same as the $g^3$ terms of
$\beta_{B'B}^{i,\Pi}$.

\section{Matrix Elements of $\gamma_{B'}^\Pi (B)$}

In this Appendix we give the matrix elements of $\gamma_{B'}^\Pi (B)$.
\begin{eqnarray}
\begin{array}{ll}
\displaystyle
\gamma_N^\pi (N) = \frac{1}{4 N_c^2} [ N_c (N_c + 2) g + 3h ]^2, \qquad &
\displaystyle
\gamma_\Delta^\pi (N) = \frac12 (N_c-1)(N_c+5) g^2, \\[3mm]
\displaystyle
\gamma_\Lambda^K (N) = \frac{3(N_c+3)}{8N_c^2} [ N_c \, g + h ]^2, &
\displaystyle
\gamma_\Sigma^K (N) = \frac{N_c-1}{8N_c^2} [ N_c \, g - 3h ]^2, \\[3mm]
\displaystyle
\gamma_{\Sigma^*}^K (N) = (N_c-1) g^2, &
\displaystyle
\gamma_N^\eta (N) = \frac14 [ g + h ]^2, \\[3mm]
\displaystyle
\gamma_\Delta^\eta (N) = 0. &  \\
\end{array}
\end{eqnarray}

\begin{eqnarray}
\begin{array}{ll}
\displaystyle
\gamma_\Lambda^\pi (\Lambda) = 0, &
\displaystyle
\gamma_\Sigma^\pi (\Lambda) = \frac14 (N_c-1)(N_c+3) g^2, \\[3mm]
\displaystyle
\gamma_{\Sigma^*}^\pi (\Lambda) = \frac12 (N_c-1)(N_c+3) g^2, \qquad &
\displaystyle
\gamma_N^K (\Lambda) = \frac{3(N_c+3)}{4N_c^2} [ N_c \, g + h ]^2, \\[3mm]
\displaystyle
\gamma_\Delta^K (\Lambda) = 0, &
\displaystyle
\gamma_\Xi^K (\Lambda) = \frac{N_c-1}{4N_c^2} [ N_c \, g + 3h ]^2, \\[3mm]
\displaystyle
\gamma_{\Xi^*}^K (\Lambda) = 2(N_c-1)g^2 , &
\displaystyle
\gamma_\Lambda^\eta (\Lambda) = \frac{1}{4N_c^2} [ 2N_c \, g - (N_c-3)h ]^2,
\\[3mm]
\displaystyle
\gamma_\Sigma^\eta (\Lambda) = 0 &
\displaystyle
\gamma_{\Sigma^*}^\eta (\Lambda) = 0.
\end{array}
\end{eqnarray}

\begin{eqnarray}
\begin{array}{ll}
\displaystyle
\gamma_\Lambda^\pi (\Sigma) = \frac{1}{12} (N_c-1)(N_c+3) g^2, &
\displaystyle
\gamma_\Sigma^\pi (\Sigma) = \frac{1}{6N_c^2} [ N_c(N_c+1)g + 6h ]^2,
\\[3mm]
\displaystyle
\gamma_{\Sigma^*}^\pi (\Sigma) = \frac{1}{12} (N_c+1)^2 g^2, &
\displaystyle
\gamma_N^K (\Sigma) = \frac{N_c-1}{12 N_c^2} [ N_c \, g - 3h ]^2, \\[3mm]
\displaystyle
\gamma_\Delta^K (\Sigma) = \frac23 (N_c+5) g^2, &
\displaystyle
\gamma_\Xi^K (\Sigma) = \frac{N_c+3}{36 N_c^2} [ 5N_c \, g + 3 h ]^2,
\\[3mm]
\displaystyle
\gamma_{\Xi^*}^K (\Sigma) = \frac29 (N_c+3) g^2, &
\displaystyle
\gamma_\Lambda^\eta (\Sigma) = 0, \\[3mm]
\displaystyle
\gamma_\Sigma^\eta (\Sigma) =
\frac{1}{4N_c^2} [ 2N_c \, g + (N_c-3) h ]^2, \qquad &
\displaystyle
\gamma_{\Sigma^*}^\eta (\Sigma) = 2g^2.
\end{array}
\end{eqnarray}

\begin{eqnarray}
\begin{array}{ll}
\displaystyle
\gamma_\Xi^\pi (\Xi) = \frac{1}{36 N_c^2} [ N_c^2 g - 9h ]^2, &
\displaystyle
\gamma_{\Xi^*}^\pi (\Xi) = \frac29 N_c^2 g^2, \\[3mm]
\displaystyle
\gamma_\Lambda^K (\Xi) = \frac{N_c-1}{8 N_c^2} [ N_c \, g + 3h ]^2, &
\displaystyle
\gamma_\Sigma^K (\Xi) = \frac{N_c+3}{24 N_c^2} [ 5N_c \, g + 3h ]^2, \\[3mm]
\displaystyle
\gamma_{\Sigma^*}^K (\Xi) = \frac13 (N_c+3) g^2, &
\displaystyle
\gamma_\Omega^K (\Xi) = (N_c+1) g^2, \\[3mm]
\displaystyle
\gamma_\Xi^\eta (\Xi) =
\frac{1}{4 N_c^2} [ 3N_c \, g - (N_c-6)h ]^2, \qquad &
\displaystyle
\gamma_{\Xi^*}^\eta (\Xi) = 2g^2.
\end{array}
\end{eqnarray}

\begin{eqnarray}
\begin{array}{ll}
\displaystyle
\gamma_N^\pi (\Delta) = \frac18 (N_c-1)(N_c+5) g^2, &
\displaystyle
\gamma_\Delta^\pi (\Delta) = \frac{1}{4 N_c^2} [ N_c(N_c+2)g + 15 h ]^2,
\\[3mm]
\displaystyle
\gamma_\Lambda^K (\Delta) = 0, &
\displaystyle
\gamma_\Sigma^K (\Delta) = \frac14 (N_c+5) g^2, \\[3mm]
\displaystyle
\gamma_{\Sigma^*}^K (\Delta) =
\frac{5(N_c+5)}{16N_c^2} [ N_c \, g + 3h ]^2, \qquad &
\displaystyle
\gamma_N^\eta (\Delta) = 0, \\[3mm]
\displaystyle
\gamma_\Delta^\eta (\Delta) = \frac54 [ g+h ]^2. & 
\end{array}
\end{eqnarray}

\begin{eqnarray}
\begin{array}{ll}
\displaystyle
\gamma_\Lambda^\pi (\Sigma^*) = \frac{1}{12} (N_c-1)(N_c+3) g^2, &
\displaystyle
\gamma_\Sigma^\pi (\Sigma^*) = \frac{1}{24} (N_c+1)^2 g^2, \\[3mm]
\displaystyle
\gamma_{\Sigma^*}^\pi (\Sigma^*) =
\frac{5}{24N_c^2} [ N_c(N_c+1)g + 12 h ]^2, \qquad &
\displaystyle
\gamma_N^K (\Sigma^*) = \frac13 (N_c-1) g^2, \\[3mm]
\displaystyle
\gamma_\Delta^K (\Sigma^*) = \frac{5(N_c+5)}{12 N_c^2} [ N_c \, g + 3h ]^2, &
\displaystyle
\gamma_\Xi^K (\Sigma^*) = \frac19 (N_c+3) g^2, \\[3mm]
\displaystyle
\gamma_{\Xi^*}^K (\Sigma^*) = \frac{5(N_c+3)}{9N_c^2} [ N_c \, g + 3h ]^2, &
\displaystyle
\gamma_\Lambda^\eta (\Sigma^*) = 0, \\[3mm]
\displaystyle
\gamma_\Sigma^\eta (\Sigma^*) = g^2, &
\displaystyle
\gamma_{\Sigma^*}^\eta (\Sigma^*) = \frac{5(N_c-3)^2}{4N_c^2} h^2.
\end{array}
\end{eqnarray}

\begin{eqnarray}
\begin{array}{ll}
\displaystyle
\gamma_\Xi^\pi (\Xi^*) = \frac19 N_c^2 g^2, &
\displaystyle
\gamma_{\Xi^*}^\pi (\Xi^*) = \frac{5}{36 N_c^2} [ N_c^2 g + 9h ]^2,
\\[3mm]
\displaystyle
\gamma_\Lambda^K (\Xi^*) = \frac12 (N_c-1) g^2, &
\displaystyle
\gamma_{\Sigma}^K (\Xi^*) = \frac16 (N_c+3) g^2, \\[3mm]
\displaystyle
\gamma_{\Sigma^*}^K (\Xi^*) =
\frac{5(N_c+3)}{6N_c^2} [ N_c \, g + 3h ]^2, \qquad &
\displaystyle
\gamma_\Omega^K (\Xi^*) =
\frac{5(N_c+1)}{8 N_c^2} [ N_c \, g + 3h ]^2, \\[3mm]
\displaystyle
\gamma_\Xi^\eta (\Xi^*) = g^2, &
\displaystyle
\gamma_{\Xi^*}^\eta (\Xi^*) = \frac{5}{4 N_c^2} [ N_c \, g - (N_c-6) h ]^2.
\end{array}
\end{eqnarray}

\begin{eqnarray}
\begin{array}{ll}
\displaystyle
\gamma_\Omega^\pi (\Omega) = 0, &
\displaystyle
\gamma_\Xi^K (\Omega) = (N_c+1) g^2, \\[3mm]
\displaystyle
\gamma_{\Xi^*}^K (\Omega) =
\frac{5(N_c+1)}{4 N_c^2} [ N_c \, g + 3h ]^2, \qquad &
\displaystyle
\gamma_\Omega^\eta (\Omega) = \frac{5}{4 N_c^2} [ 2N_c \, g - (N_c-9) h ]^2.
\end{array}
\end{eqnarray}

For the fictitious intermediate states, we have

\begin{eqnarray}
\begin{array}{ll}
\displaystyle
\gamma_{\tilde \Sigma^*}^\pi (\Sigma) =
\frac{5}{12} (N_c-3)(N_c+5) g^2, \qquad &
\displaystyle
\gamma_{\tilde \Xi}^K (\Sigma) =
\frac{2(N_c-3)}{9 N_c^2} [ N_c \, g - 3h ]^2, \\[3mm]
\displaystyle
\gamma_{\tilde \Xi^*}^K (\Sigma) =
\frac{10}{9} (N_c-3) g^2.
\end{array}
\end{eqnarray}

\begin{eqnarray}
\begin{array}{ll}
\displaystyle
\gamma_{\tilde \Xi}^\pi (\Xi) =
\frac{2}{9} (N_c-3)(N_c+3) g^2, \qquad &
\displaystyle
\gamma_{\tilde \Xi^*}^\pi (\Xi) =
\frac{5}{18} (N_c-3)(N_c+3) g^2, \\[3mm]
\displaystyle
\gamma_{\tilde \Omega}^K (\Xi) =
\frac{(N_c-3)}{3 N_c^2} [ N_c \, g + 3h ]^2, \qquad &
\displaystyle
\gamma_{\tilde \Omega^*}^K (\Xi) =
\frac{15}{9} (N_c-3) g^2 .
\end{array}
\end{eqnarray}

\begin{eqnarray}
\begin{array}{ll}
\displaystyle
\gamma_{\tilde \Delta^{**}}^\pi (\Delta) =
\frac{3}{8} (N_c-3)(N_c+7) g^2, \qquad &
\displaystyle
\gamma_{\tilde \Sigma^{*}}^K (\Delta) =
\frac{3(N_c-3)}{16 N_c^2} [ N_c \, g - 5h ]^2, \\[3mm]
\displaystyle
\gamma_{\tilde \Sigma^{**}}^K (\Delta) =
\frac{3}{4} (N_c-3) g^2.
\end{array}
\end{eqnarray}

\begin{eqnarray}
\begin{array}{ll}
\displaystyle
\gamma_{\tilde \Sigma^{*}}^\pi (\Sigma^*) =
\frac{1}{24} (N_c-3)(N_c+5) g^2, \qquad &
\displaystyle
\gamma_{\tilde \Sigma^{**}}^\pi (\Sigma^*) =
\frac{3}{8} (N_c-3)(N_c+5) g^2, \\[3mm]
\displaystyle
\gamma_{\tilde \Xi}^K (\Sigma^*) =
\frac{1}{18} (N_c-3) g^2, \qquad &
\displaystyle
\gamma_{\tilde \Xi^*}^K (\Sigma^*) =
\frac{(N_c-3)}{36 N_c^2} [ N_c \, g - 15h]^2, \\[3mm]
\displaystyle
\gamma_{\tilde \Xi^{**}}^K (\Sigma^*) =
\frac{3}{2} (N_c-3) g^2.
\end{array}
\end{eqnarray}

\begin{eqnarray}
\begin{array}{ll}
\displaystyle
\gamma_{\tilde \Xi}^\pi (\Xi^*) =
\frac{1}{72} (N_c-3)(N_c+3) g^2, \qquad &
\displaystyle
\gamma_{\tilde \Xi^{*}}^\pi (\Xi^*) =
\frac{1}{9} (N_c-3)(N_c+3) g^2, \\[3mm]
\displaystyle
\gamma_{\tilde \Xi^{**}}^\pi (\Xi^*) =
\frac{3}{8} (N_c-3)(N_c+3) g^2, \qquad &
\displaystyle
\gamma_{\tilde \Omega}^K (\Xi^*) =
\frac{1}{12} (N_c-3) g^2, \\[3mm]
\displaystyle
\gamma_{\tilde \Omega^{*}}^K (\Xi^*) =
\frac{(N_c-3)}{24 N_c^2} [ N_c \, g + 15h ]^2, \qquad &
\displaystyle
\gamma_{\tilde \Omega^{**}}^K (\Xi^*) =
\frac{9}{4} (N_c-3) g^2.
\end{array}
\end{eqnarray}

\begin{eqnarray}
\begin{array}{ll}
\displaystyle
\gamma_{\tilde \Omega}^\pi (\Omega) =
\frac{1}{8} (N_c-3)(N_c+1) g^2, \qquad &
\displaystyle
\gamma_{\tilde \Omega^*}^\pi (\Omega) =
\frac{1}{4} (N_c-3)(N_c+1) g^2, \\[3mm]
\displaystyle
\gamma_{\tilde \Omega^{**}}^\pi (\Omega) =
\frac{3}{8} (N_c-3)(N_c+1) g^2, \qquad &
\displaystyle
\gamma_{\tilde {\rm S}^{*}}^K (\Omega) =
\frac{3(N_c-3)}{4 N_c^2} [ N_c \, g + 5h ]^2, \\[3mm]
\displaystyle
\gamma_{\tilde {\rm S}^{**}}^K (\Omega) = 3(N_c-3) g^2,
\end{array}
\end{eqnarray}
and the others are zero. Note that all the matrix elements with fictitious
intermediate state contain the factor $(N_c-3)$ so that they vanish
in the real world with $N_c=3$.



\renewcommand{\arraystretch}{1.1}
\begin{table}
\centering
\begin{tabular}{c|r|r|r|r|r|r}
  & $\{ {\cal U} \}$ & 
$\{ \sigma_j \} \{ \sigma_j \}$ & $\{ {\cal S} \}$ &
$\{ {\cal S} \} \{ {\cal S} \}$ & 
$\{ {\cal S} \sigma_j \} \{ \sigma_j \}$ &
$\{ {\cal U} \sigma_j \} \{ {\cal S} \sigma_j \}$ \\ \hline
$N$       & $2n+1$ &  3 & 0 & 0 & 0   & 0 \\
$\Lambda$ & $2n$   &  3 & 1 & 1 & 3   & 0 \\
$\Sigma$  & $2n$   &  3 & 1 & 1 & $-1$& $-4$ \\
$\Xi$     & $2n-1$ &  3 & 2 & 4 & 4   & $-4$ \\ \hline
$\Delta$  & $2n+1$ & 15 & 0 & 0 & 0   & 0 \\
$\Sigma^*$& $2n$   & 15 & 1 & 1 & 5   & 2 \\
$\Xi^*$   & $2n-1$ & 15 & 2 & 4 & 10  & 2 \\
$\Omega$  & $2n-2$ & 15 & 3 & 9 & 15  & 0 \\
\hline \hline
$\tilde \Xi$         & $2n-1$ & $ 3$ & $2$ & $ 4$ & $-2$ & $-10$ \\
$\tilde \Omega$      & $2n-2$ & $ 3$ & $3$ & $ 9$ & $ 5$ & $-10$ \\ \hline
$\tilde \Sigma^*$    & $2n$   & $15$ & $1$ & $ 1$ & $-3$ & $-6$ \\
$\tilde \Xi^*$       & $2n-1$ & $15$ & $2$ & $ 4$ & $ 4$ & $-4$ \\
$\tilde \Omega^*$    & $2n-2$ & $15$ & $3$ & $ 9$ & $11$ & $-4$ \\
$\tilde {\rm S}^*$   & $2n-3$ & $15$ & $4$ & $16$ & $18$ & $-6$ \\ \hline
$\tilde \Delta^{**}$ & $2n+1$ & $35$ & $0$ & $ 0$ & $ 0$ & $ 0$ \\
$\tilde \Sigma^{**}$ & $2n$   & $35$ & $1$ & $ 1$ & $ 7$ & $ 4$ \\
$\tilde \Xi^{**}$    & $2n-1$ & $35$ & $2$ & $ 4$ & $14$ & $ 6$ \\
$\tilde \Omega^{**}$ & $2n-2$ & $35$ & $3$ & $ 9$ & $21$ & $ 6$ \\
$\tilde {\rm S}^{**}$& $2n-3$ & $35$ & $4$ & $16$ & $28$ & $ 4$ \\
\end{tabular}
\bigskip
\caption{Matrix elements of various operators for baryon states}
\label{tab:me1}
\end{table}

\renewcommand{\arraystretch}{1.0}
\begin{table}
\centering
\begin{tabular}{c|r|r|r|r|r}
Particle & $O(\varepsilon^1)$ & $O(\varepsilon^2)$ & $O(\varepsilon^{3,4})$ &
$O(\varepsilon^{5})$ & Expt. \\ \hline
$N$        & 1142 &  982 &  939 &  937 &  939 \\
$\Lambda$  & 1142 & 1141 & 1117 & 1119 & 1116 \\
$\Sigma$   & 1142 & 1141 & 1183 & 1183 & 1193 \\
$\Xi$      & 1142 & 1300 & 1328 & 1327 & 1318 \\ \hline
$\Delta$   & 1456 & 1217 & 1238 & 1236 & 1232 \\
$\Sigma^*$ & 1456 & 1376 & 1383 & 1386 & 1385 \\
$\Xi^*$    & 1456 & 1535 & 1528 & 1530 & 1530 \\
$\Omega$   & 1456 & 1694 & 1673 & 1670 & 1672 \\ \hline
$\sqrt{\chi^2}$ & 424 & 79 & 16 & 15 & \\ \hline
$a$  & 1063.0 & 923.9 & 863.7   & 862.4 \\
$b$  & 26.2   & 19.5  & 25.0    & 24.9  \\
$c$  &  ---   & 159.0 & 227.8   & 96.5  \\
$d$  &  ---   &  ---  & $-16.6$ & 51.8 \\
$e$  & ---    &  ---  &  ---    & $-70.4$ \\
$f$  &  ---   &  ---  &  ---    & $-67.8$
\end{tabular}
\bigskip
\caption{Best fit of baryon masses (tree) in the unit of MeV at each order
of $\varepsilon$ using the formula (\protect\ref{massform}).}
\label{tab:fit}
\end{table}

\begin{table}
\centering
\begin{tabular}{c|ccc} 
operator        & $g^2$ term & $gh$ term & $h^2$ term \\ \hline
${\cal O}^{\pi}$  & $N_c^2$      & $N_c^0$     & $N_c^{-2}$ \\
${\cal O}^{K}$    & $N_c^1$      & $N_c^0$     & $N_c^{-1}$ \\
${\cal O}^{\eta}$ & $N_c^0$      & $N_c^0$     & $N_c^{0}$
\end{tabular}
\bigskip
\caption{The leading order of operator ${\cal O}^\Pi$ depending on the
coupling constants.}
\label{taborder}
\end{table}

\end{document}